\documentclass{aa}

\usepackage[svgnames]{xcolor}
\usepackage{graphicx}
\usepackage{breqn} 
\graphicspath{ {images/} }

\usepackage{txfonts}
\usepackage{subcaption}
\usepackage{booktabs}
\usepackage{threeparttable}
\usepackage{float}
\usepackage{color}
\usepackage{colortbl} 
\usepackage{amsmath}
\usepackage{epstopdf}
\usepackage{enumitem}
\usepackage{multicol}
\usepackage{multirow}
\usepackage{gensymb}
\usepackage{chngcntr}
\usepackage{ulem}
\usepackage{rotating}
\usepackage{dblfloatfix} 
\usepackage{dblfloatfix} 
\usepackage[export]{adjustbox}
\usepackage{ulem,color}

\usepackage{natbib}
\usepackage[colorlinks, citecolor={blue}]{hyperref}
\def\Beq{\mbox{\,$B_{\rm eq}$\,}}

\def\gsim{\!\!\!\phantom{\ge}\smash{\buildrel{}\over
{\lower2.5dd\hbox{$\buildrel{\lower2dd\hbox{$\displaystyle>$}}\over
                               \sim$}}}\,\,}

\def\asec{\ifmmode ^{\prime\prime}\else$^{\prime\prime}$\fi}

\def\Msunyr{\mbox{\,${\rm M_{\odot}\, yr^{-1}}$}}

\def\degs{\ifmmode ^{\circ}\else$^{\circ}$\fi}
\def\amin{\ifmmode ^{\prime}\else$^{\prime}$\fi}
\def\asec{\ifmmode ^{\prime\prime}\else$^{\prime\prime}$\fi}
\def\arcmin{\ifmmode ^{\prime}\else$^{\prime}$\fi}
\def\arcsec{\ifmmode ^{\prime\prime}\else$^{\prime\prime}$\fi}

\def\degs{\ifmmode ^{\circ}\else$^{\circ}$\fi}
\def\amin{\ifmmode ^{\prime}\else$^{\prime}$\fi}

\def\cm{\mbox{\,cm}}

\def\cm3{\mbox{\,cm$^{-3}$}}

\def\lsim{\!\!\!\phantom{\le}\smash{\buildrel{}\over
 {\lower2.5dd\hbox{$\buildrel{\lower2dd\hbox{$\displaystyle<$}}\over
                                 \sim$}}}\,\,}
\def\gsim{\!\!\!\phantom{\ge}\smash{\buildrel{}\over
{\lower2.5dd\hbox{$\buildrel{\lower2dd\hbox{$\displaystyle>$}}\over
                               \sim$}}}\,\,}

\def\ARP299{$\rm Arp\, 299$}
\def\fixedlabel#1#2{%
  \@bsphack%
  \protected@write\@auxout{}%
         {\string\newlabel{#1}{{#2}{\thepage}}}%
  \@esphack}
\makeatother

\begin{document} 

   \title{Subarcsecond LOFAR imaging of Arp299 at 150 MHz}
   \subtitle{Tracing the nuclear and diffuse extended emission of a bright LIRG}
   \titlerunning{Subarcsecond LOFAR imaging of Arp299 at 150 MHz}

   \author{N. Ram\'irez-Olivencia \inst{1},
          E. Varenius \inst{2},
          M. P\'erez-Torres \inst{1},
          A. Alberdi \inst{1}, 
          J. E. Conway \inst{2},
          A. Alonso-Herrero \inst{3},
          M. Pereira-Santaella \inst{4},
          R. Herrero-Illana \inst{5,6}
          }
\authorrunning{N. Ram\'irez-Olivencia et al.}
   \institute{{$^1$} Instituto de Astrof\'isica de Andaluc\'ia (IAA-CSIC),
              Glorieta de la Astronom\'ia s/n, 18008 Granada, Spain,
              \email{naim.ram.oli@gmail.com}\\
              {$^2$} Department of Space, Earth and Environment, Chalmers University of Technology, Onsala Space Observatory, 439 92 Onsala, Sweden \\
              {$^3$} Centro de Astrobiología (CAB, CSIC-INTA), ESAC Campus, E-28692 Villanueva de la Cañada, Madrid, Spain \\
              {$^4$} Centro de Astrobiología (CAB, CSIC-INTA), Carretera de Torrejón a Ajalvir, E-28880 Torrejón de Ardoz, Madrid, Spain \\
              {$^5$} European Southern Observatory (ESO), Alonso de C\'ordova 3107, Vitacura, Casilla 19001, Santiago de Chile, Chile \\
              {$^6$} Institute of Space Sciences (ICE, CSIC), Campus UAB, Carrer de Magrans, E-08193 Barcelona, Spain \\
            }


 
  \abstract
   {Arp~299 is the brightest luminous infrared galaxy (LIRG) within 50 Mpc, with  IR luminosity  log(L$_{IR}/L_\odot$)=11.9.  It provides a unique laboratory for testing physical processes in merging galaxies. 
  }  
   {We study for the first time the low-frequency ($\sim$150 MHz) radio brightness distribution of Arp~299 at subarcsecond resolution, tracing in both compact and extended emission regions the local spectral energy distribution (SED) in order to characterize the dominant emission and absorption processes. 
   }
   {We analysed the spatially resolved emission of Arp 299 revealed by 150 MHz international baseline  Low-Frequency Array (LOFAR) and 1.4, 5.0, and 8.4 GHz Very Large Array (VLA) observations.  
   }
   {We present the first subarcsecond (0.4"$\sim$100~pc) image of the whole Arp~299 system at 150~MHz. The high surface brightness sensitivity of our LOFAR observations 
   ($\sim$100 $\mu$Jy/beam) allowed us to detect all of the nuclear components detected at higher frequencies, as well as the extended steep-spectrum emission surrounding the nuclei. We obtained spatially resolved, two-point spectral index maps for the whole galaxy: the compact nuclei show relatively flat spectra, while the extended, diffuse component shows a steep spectrum. 
   We fitted the radio  SED of the nuclear regions using two different models: a continuous free-free medium model and a clumpy model. The continuous model can explain the SED of the nuclei assuming a population of relativistic electrons subjected to synchrotron, bremsstrahlung, and ionization losses.
   The clumpy model fits assuming relativistic electrons with  negligible energy losses, and thermal fractions that are more typical of star-forming galaxies than those required for the continuous model. 
   }   
   {Our results confirm the usefulness of combining spatially resolved radio imaging at both MHz and GHz frequencies to characterize in detail the radio emission properties of LIRGs from the central 100 pc out to the kiloparsec galaxy-wide scales. 
  }

   \maketitle


\section{Introduction}\label{sec:introduction}

Luminous and Ultraluminous infrared galaxies (LIRGs and ULIRGs, respectively) are the most luminous (log($\rm L _{IR}/L_{\odot}) \geq 11.0$ for LIRGs, log($\rm L_{IR}/L_{\odot}) \geq 12$ for ULIRGs \citep{sanders}) galaxies in the infrared domain, with $L_{IR}$ making up the bulk of their total bolometric luminosity. The number of ULIRGs increases with redshift \citep{magnelli}, suggesting that they were dominant in the infrared sky at z=1-2. The multifrequency study of their characteristics and behaviour in the local Universe can serve as a template to understand the characteristics of ULIRGs in the high-$z$ universe.

 Arp~299 (also known as NGC~3690-A~+~NGC~3690-B, Mrk~171, VV~118, IRAS~11257+5850) at 
  $D=44.8$~Mpc is the brightest LIRG within 50 Mpc, and with 
  $L_{\rm IR} \approx 8\times10^{11}$ L$_{\odot}$
  it approaches the ULIRG class. The Arp~299 system consists of a pair of two galaxies in an early major merging process \citep{keel}, with three well separated regions (\citealt{gehrz,neff}). Region A is the nucleus of the eastern galaxy, B is the nucleus of the western galaxy, and C and C' are off-nuclear  star-forming regions in the overlap of the two disk components of the system \citep{telesco}. Both nuclei A and B are known to host AGNs: Arp~299-A hosts a low-luminosity AGN (LLAGN;\citealt{perez-torres2010}), while Arp~299-B  hosts an obscured Seyfert-like AGN \citep{della,garcia-marin,alonso-herrero2013}.

\begin{table*}
\caption{\label{tab:general_results}Summary of  LOFAR and JVLA observations and their different configurations.}
\begin{tabular}{lllllccc}
\hline
       Telescope & Frequency & Project & \begin{tabular}[c]{@{}l@{}}Observing Date\\ DD-MM-YYY\end{tabular} & \begin{tabular}[c]{@{}l@{}}rms\\ {[}$\mu$Jy/beam{]}\end{tabular} & \begin{tabular}[c]{@{}c@{}}Peak\\ {[}mJy/beam{]}\end{tabular} & \begin{tabular}[c]{@{}c@{}}Integrated Flux\\ {[}Jy{]}\end{tabular} & \begin{tabular}[c]{@{}c@{}}Beam\\ (bmaj(")$\times$bmin("),bpa($^{\circ}$)))\end{tabular} \\ \hline
LOFAR  & 150 MHz   & LC5-020 & 22-02-2016                                                         & 98                                                             & 23.2                                                          & 0.90$\pm$0.13                                                              & 0.44$\times$0.42,-52                                                                \\
JVLA-L & 1.4 GHz   & GP053   & 12-06-2015                                                         & 59                                                             & 113.1                                                         & 0.59$\pm$0.08                                                               & 1.35$\times$1.11,81                                                                \\
JVLA-C & 5.0 GHz   & GP053   & 20-10-2014                                                         & 18                                                             & 70.6                                                          & 0.35$\pm$0.05                                                               & 0.79$\times$0.48,77                                                                 \\
JVLA-X & 8.4 GHz   & GP053   & 03-11-2014                                                         & 11                                                             & 44.6                                                          & 0.43$\pm$0.06                                                              & 0.54$\times$0.36,86                                                                \\ \hline
\label{tab:telescopes}
\end{tabular}
\end{table*}

 The inner kiloparsec region of Arp~299 harbours a large amount of molecular gas in its circumnuclear regions ($M({\rm H}_2) =7.5 \times 10^9$ M$_{\odot}$; \citealt{aalto97}), and radio observations with high angular resolution have unveiled an intense star formation burst in their nuclear regions, within the central 100 pc \citep{perez-torres2009,bondi12,romero-canizales1,alonso-herrero2000}.
 The energy released by the compact starburst in the central region of Arp~299A is so large as to power a large-scale superwind, or outflow, as recently unveiled by \citet{ramirez-olivencia} using the sub-mJy sensitivity and  subarcsecond angular resolution capabilities of LOFAR at 150~MHz (LOFAR; \citealt{LOFAR2013}).

The high sensitivity and angular resolution of LOFAR allow detailed studies of the very low-frequency (150 MHz corresponds to a wavelength of almost 2 metres) emission and absorption properties of nearby galaxies \citetext{e.g. M82 \citealp{varenius2015} and Arp~220 \citealp{varenius2016}}. Together with matched-resolution images at higher frequencies, the spatially resolved spectral energy distribution (SED) allows us to disentangle regions with different (thermal and/or non-thermal) spectral properties. For example, synchrotron emission in the subgigahertz regime can suffer significant free-free absorption by H~II regions \citep[e.g.][]{Condon92}. Since H~II regions are found within star-forming regions, the radio emission arising from young massive stars and supernovae are expected to be partially suppressed at megahertz frequencies due to the free electrons pervading the H~II regions. This absorption can be traced well with LOFAR observations.

In this paper we present the first subarcsecond angular resolution images of the whole Arp~299 system at 150 MHz. We also present a combined interpretation of the LOFAR observations at 150 MHz and JVLA observations at 1.4, 5.0, and 8.4GHz (L-band, C-band, and X-band, respectively).


\section{Observations and data reduction} \label{sec:reduction}

\subsection{LOFAR at 150~MHz}

We  observed Arp~299 with LOFAR between 22 and 23 February 2016 (project code LC5\_20; P.I. P\'erez-Torres, see Table\ref{tab:general_results}), for a total of 12~h, including the international stations. We centred our observations at a frequency of 150~MHz for three simultaneous beams with the same 32~MHz coverage. We used the nearby compact source J1127+5841 for phase calibration, J1128+5925 for delay/rate/amplitude/bandpass calibration, and  3C~295 for absolute flux density calibration.  Further details of the LOFAR observations and data reduction are given in  \citet{ramirez-olivencia}.

\subsection{JVLA at 1.4~GHz, 5~GHz, and 8.4~GHz}

We show in Table \ref{tab:general_results} the summary for the observations taken at 1.4~GHz, 5.0~GHz, and 8.4~GHz with the JVLA taken in October 2014  (5.0 GHz, C-configuration), November 2014 (8.4 GHz, C-configuration), and June 2015 (1.4 GHz, BnA to A configuration) under the global VLBI experiment GP053 (P.I.: P\'erez-Torres).

We averaged the data to 10~sec in time and 2~MHz per channel before calibration to reduce the processing time. We then used AOFlagger v2.8 to remove radio frequency interference (RFI), and converted the CASA data measurement set into UVFITS using the CASA task \verb!exportuvfits!. We  did all calibration and imaging steps in a standard manner using AIPS release 31DEC16 and ParselTongue 2.3. We referenced the Arp~299 visibility phases   to J1128+5925, and derived the flux density scale and bandpass corrections from observations of the primary absolute flux density calibrator, 3C~286. We then imaged the Arp~299 system using the CLEAN deconvolution algorithm as implemented in the AIPS task \verb!IMAGR!. We applied a primary beam correction using the AIPS task \verb!PBCOR!.

\section{Results}

\subsection{General description of the LOFAR and JVLA images}

\begin{figure*}[!b]
    \centering
    \includegraphics[width=\textwidth]{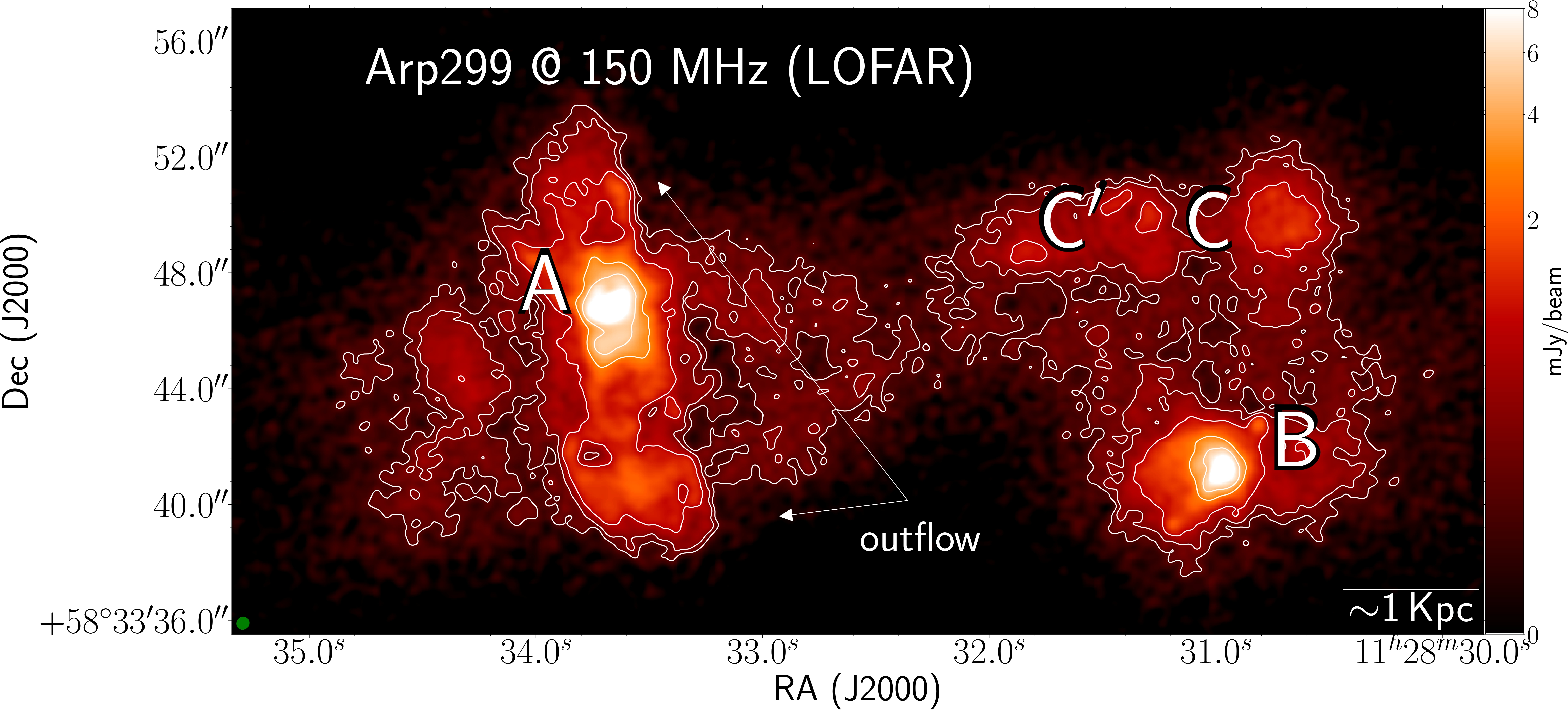}
    \caption{150~MHz LOFAR image of the LIRG Arp~299 system. Contours represent [3,5,10,20,35,50,100,200,230]$\times$rms ($\sim$98$\mu$J/beam). The classical components are labelled (A, B, C, and C$^{\prime}$), and the new outflow from A nucleus \citep{ramirez-olivencia} is also indicated. The green ellipse in the lower left corner corresponds to the synthesized beam.}
    \label{fig:LOFAR}
\end{figure*}

\begin{figure*}
\centering
\begin{subfigure}{0.5\textwidth}
  \centering
  \includegraphics[width=\textwidth]{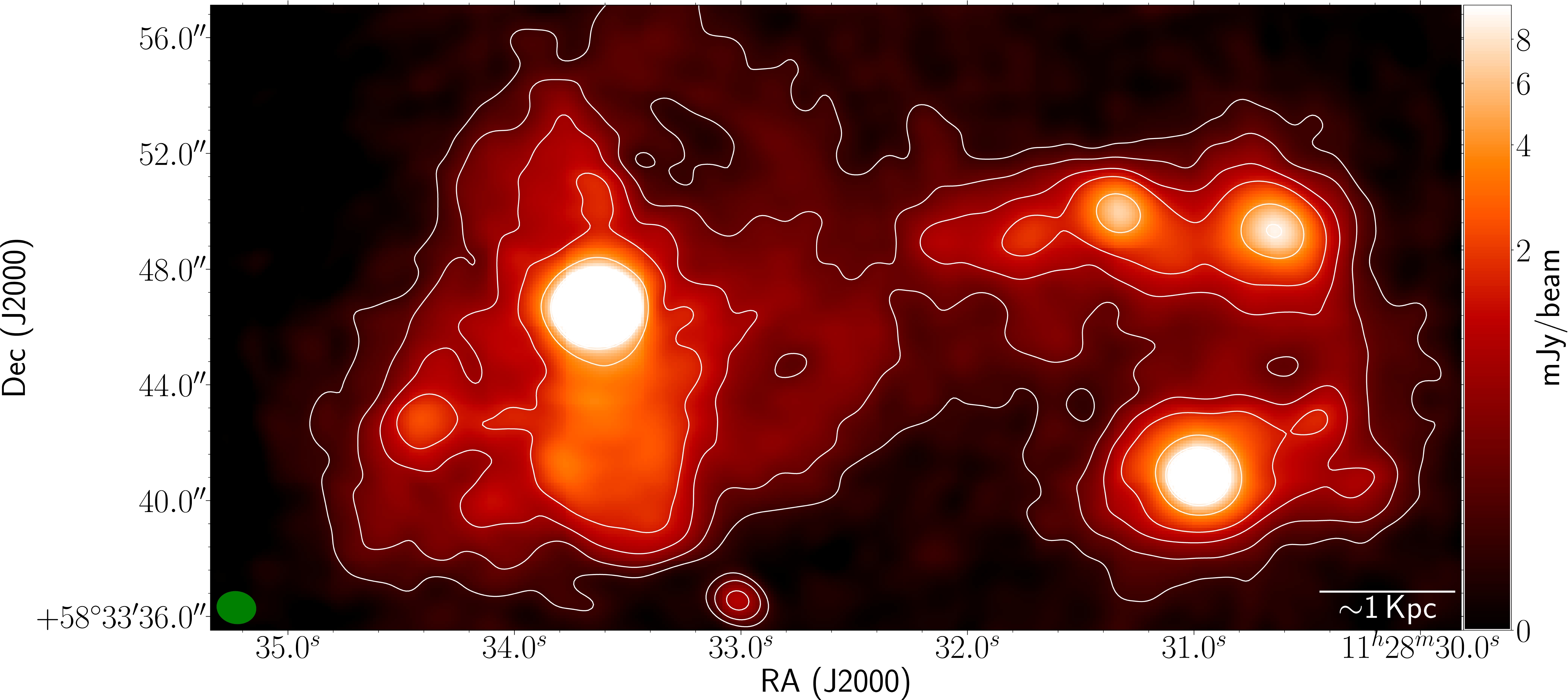}
  \label{fig:jvlal}
\end{subfigure}%
\begin{subfigure}{0.5\textwidth}
  \centering
  \includegraphics[width=\textwidth]{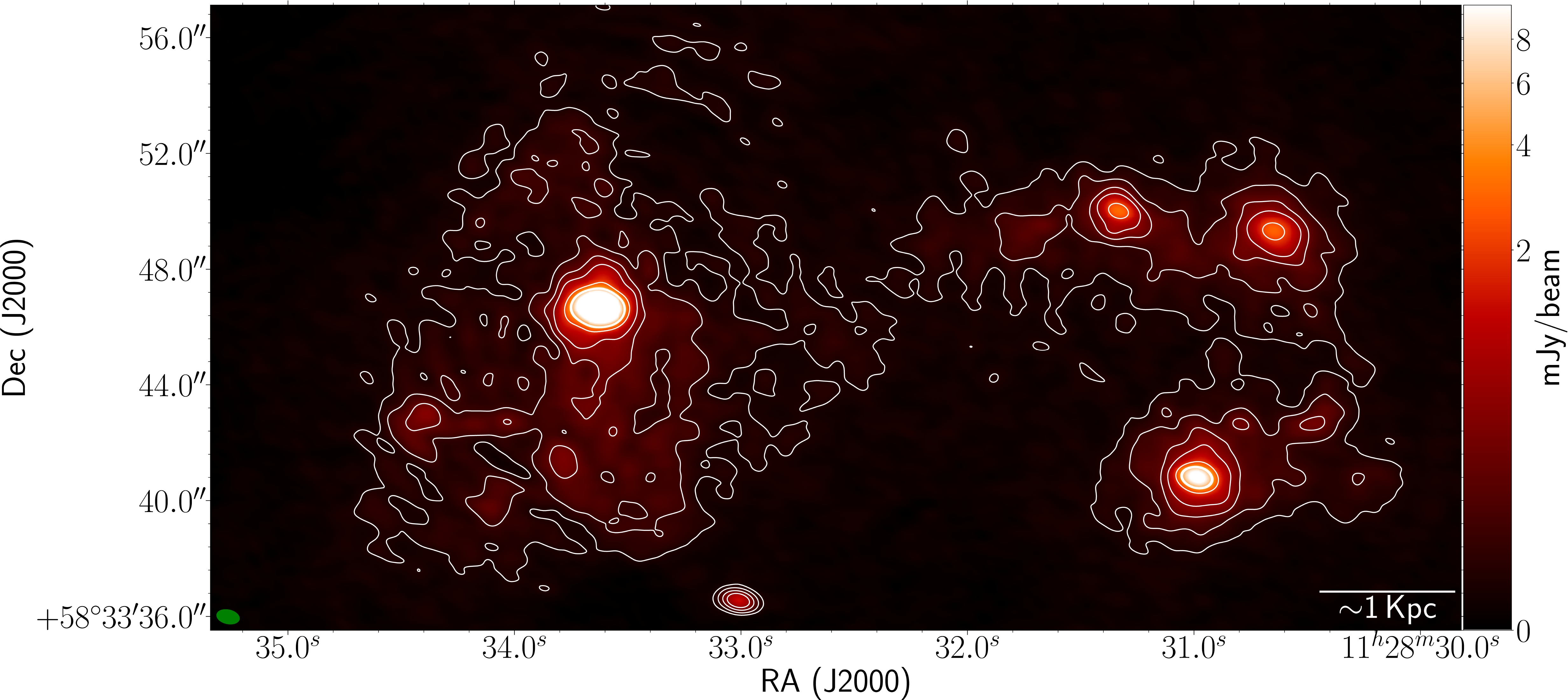}
  \label{fig:jvlac}
\end{subfigure}
\begin{subfigure}{0.5\textwidth}
  \centering
  \includegraphics[width=\textwidth]{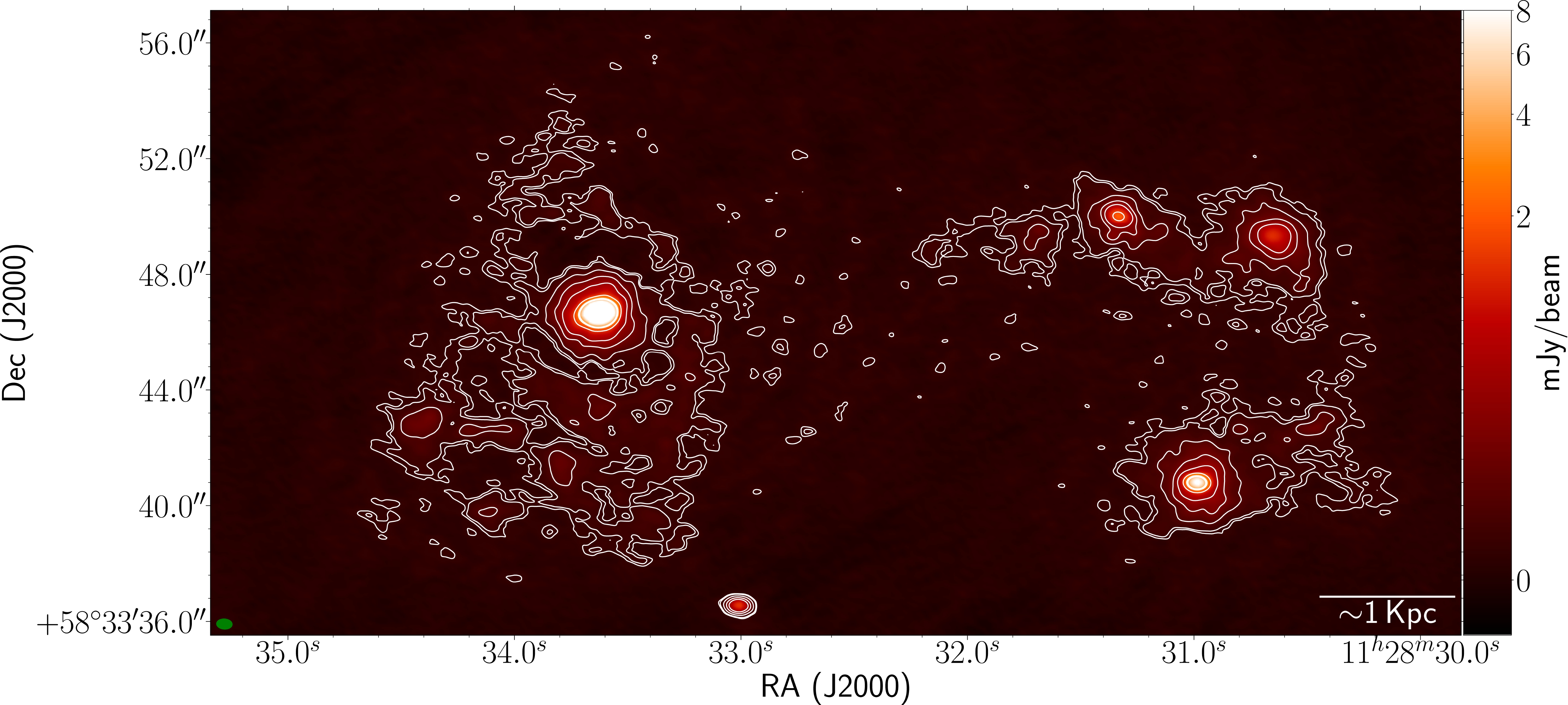}
  \label{fig:jvlax}
\end{subfigure}\\
\caption{Images of Arp~299 observed with the JVLA 
  at 1.4 GHz (top-left), 5 GHz (top-right), and 8.4 GHz (bottom).
  Contour levels are drawn at (3,5,10,20,35,50,100,200,230,250)$\times$rms at  
  each band (see Table~\ref{tab:general_results}). The green ellipse in the lower left corner corresponds to the synthesized beam.}
\label{fig:jvla}
\end{figure*}

We summarize our main results in Figs. \ref{fig:LOFAR} and \ref{fig:jvla}, which correspond to our 150 MHz (LOFAR), 1.4, 5.0, and 8.4 GHz (JVLA) images. 
The  overall morphological similarity of the JVLA images with the LOFAR image is striking, considering  the factor of $\sim$60 difference in frequency between our 150 MHz LOFAR image and the 8.4 GHz VLA image. The main components of Arp~299 are clearly detected in the LOFAR image: the two nuclei of the interacting galaxies, NGC~3690-A (where the peak of the whole LOFAR image is located) and NGC~3690-B, as well as the two prominent compact radio and IR emitting regions C and C'. We note that component D (see e.g. \citealt{neff}), which does not belong to the Arp~299 system, is not detected  in our LOFAR observations above a 3$\sigma\sim$1~mJy. The diffuse emission is also traced in detail  by the LOFAR observations, which show a clear bridge of emission that connects NGC~3690-A and NGC~3690-B. The morphological similarity of all images for the Arp~299 system, both in the compact and diffuse emission, already suggests that the main emission and absorption processes operating in this frequency regime are likely the same.

\label{sec:general}

We show in Table \ref{tab:general_results} the summary for our LOFAR and JVLA observations.  We determined the quoted total integrated flux densities as follows. 
We first measured the off-source rms and then we set up a 3$\sigma$ level in all images,
to be able to compare the emission from the nuclei and other compact features in Arp~299 at different frequencies. We then blanked out all pixels below a 3$\sigma$ threshold in the final images (not shown) and measured the total integrated flux density above this threshold. The associated uncertainties are the result of adding in quadrature the off-source rms, $\sigma_{rms}$, and the calibration uncertainty,  $\sigma_{syst,i}$, at each pixel position. We then measured $\sigma_{rms}$ in the outer regions of the system and assumed a calibration uncertainty of the flux density per pixel of 15$\%$, 10$\%$, 5$\%$, and 3$\%$ for the 150~MHz, and  1.4, 5, and 8.4~GHz observations, respectively.


\twocolumn[{%
\centering
\begin{threeparttable}
\caption{\label{tab:direct_results} Summary of  LOFAR and JVLA observations and map details. }
\begin{tabular}{llllll}
\hline
Components                 & Frequency & \begin{tabular}[c]{@{}l@{}}R.A$^{1}$.\\ (J2000)\end{tabular} & \begin{tabular}[c]{@{}l@{}}Dec.$^{1}$\\ (J2000)\end{tabular} & \begin{tabular}[c]{@{}l@{}}Peak\\ mJy/beam\end{tabular} & \begin{tabular}[c]{@{}l@{}}S$_{\nu}^{2}$\\ mJy\end{tabular} \\ \hline
\multirow{4}{*}{A nucleus} & 150~MHz & 11:28:33.661                                           & +58:33.46.840                                          & 23                                                      & 104.5$\pm$13                                            \\
                          & 1.4~GHz     & 11:28.33.63                                           & +58:33:46.66                                          & 113                                                     & 87.6$\pm$11.3                                           \\
                          & 5~GHz     & 11:28:33.63                                           & +58:33:46.65                                          & 71                                                      & 80.3$\pm$4.1                                            \\
                           & 8.4~GHz     & 11:28:33.63                                           & +58:33:46.64                                          & 44                                                      & 60.9$\pm$1.9                                            \\\hline

\multirow{4}{*}{B nucleus} & 150~MHz & 11:28:30.97                                           & +58:33.41.24                                          & 16                                                      & 36.0$\pm$9.8                                            \\
                           & 1.4~GHz     & 11:28.30.99                                           & +58:33:40.82                                          & 32                                                      & 63.4$\pm$4.3                                            \\
                           & 5~GHz     & 11:28:30.99                                           & +58:33:40.80                                          & 13                                                      & 18.5$\pm$1.3                                            \\
                           & 8.4~GHz     & 11:28:30.99                                           & +58:33:40.80                                          & 9                                                       & 13.4$\pm$0.5                                            \\\hline
\multirow{4}{*}{C compact component} & 150~MHz & 11:28:30.67                                           & +58:33.49.40                                          & 17                                                      & 25.8$\pm$4.6                                            \\
                           & 1.4~GHz     & 11:28.30.65                                           & +58:33:49.31                                          & 9                                                       & 17.9$\pm$2.3                                            \\
                           & 5~GHz     & 11:28:30.65                                           & +58:33:49.31                                          & 3                                                       & 10.5$\pm$0.6                                            \\
                           & 8.4~GHz     & 11:28:30.65                                           & +58:33:49.32                                          & 1                                                       & 8.0$\pm$0.3                                             \\\hline
\multirow{4}{*}{C$^\prime$ compact component} & 150~MHz & 11:28:31.08                                           & +58:33.51.72                                          & 14                                                      & 17.1$\pm$3.3                                            \\
                           & 1.4~GHz     & 11:28.31.34                                           & +58:33:49.99                                          & 7                                                       & 11.4$\pm$2.1                                            \\
                           & 5~GHz     & 11:28:31.33                                           & +58:33:50.01                                          & 3                                                       & 7.4$\pm$0.6                                             \\
                           & 8.4~GHz     & 11:28:31.33                                           & +58:33:50.02                                          & 1                                                       & 6.0$\pm$0.3                                             \\\hline
\end{tabular}
\begin{tablenotes}
\footnotesize
\item[1] Position of the peak for each band and compact source.\item[2] Integrated flux in the area described in Sect. \ref{sec:integrated_flux}
\end{tablenotes}

\end{threeparttable}
}]

\subsection{Spectral properties at radio wavelengths}\label{subsec:spectrum}

In this section we first present the integrated flux densities of the nuclei at different bands, and then the spectral-index maps for the whole Arp~299 system.

\subsubsection{Integrated flux density of the nuclear components}
\label{sec:integrated_flux}

We turn now to study the nuclear components A, B, C, and C' detected at all frequencies. 
In Table \ref{tab:direct_results} we show the peak and total flux density for the nuclei of Arp~299. 
We used the same physical size at each frequency for a direct comparison of our results. We first convolved the images of the nuclear components A,B, C, and C' at each frequency to the largest beam, which corresponds to that of the 1.4~GHz observations ($\theta_{\rm FWHM}\sim$1.4 arcsec), using a circular beam for simplicity. We also regridded the JVLA images to the coordinate system of LOFAR with the  CASA task \texttt{imregrid}. We then used the region delimited by the 50$\%$ of the peak of the convolved 1.4~GHz image, as a common angular size for all images.

We estimated the flux density uncertainties following the same procedure described in Sect. \ref{sec:general}. Specifically, we obtained an uncertainty map based on the total flux density map, where each pixel has a flux density uncertainty given by the sum in quadrature of the thermal noise of the image and the systematic uncertainties:  $\sqrt{\sigma_{sys}^2 + \sigma_{rms}^2}$ (see Table \ref{tab:direct_results}). 

We note that, for both   C and C', the maximum of the peaks occurs at 150~MHz, while in the case of A and B the maximum is located at 1.4~GHz. The A nucleus is the brightest component at all frequencies.  Most of the emission from the Arp~299 system comes from the extended emission, with only $\sim$20$\%$, $\sim$30$\%$, $\sim$33$\%$, and $\sim$20$\%$ at 150~MHz, and  1.4, 5, and 8.4~GHz, respectively, arising from the compact nuclear regions. 

\subsubsection{Spectral index maps}\label{sec:spix}

\begin{figure*}[hp]
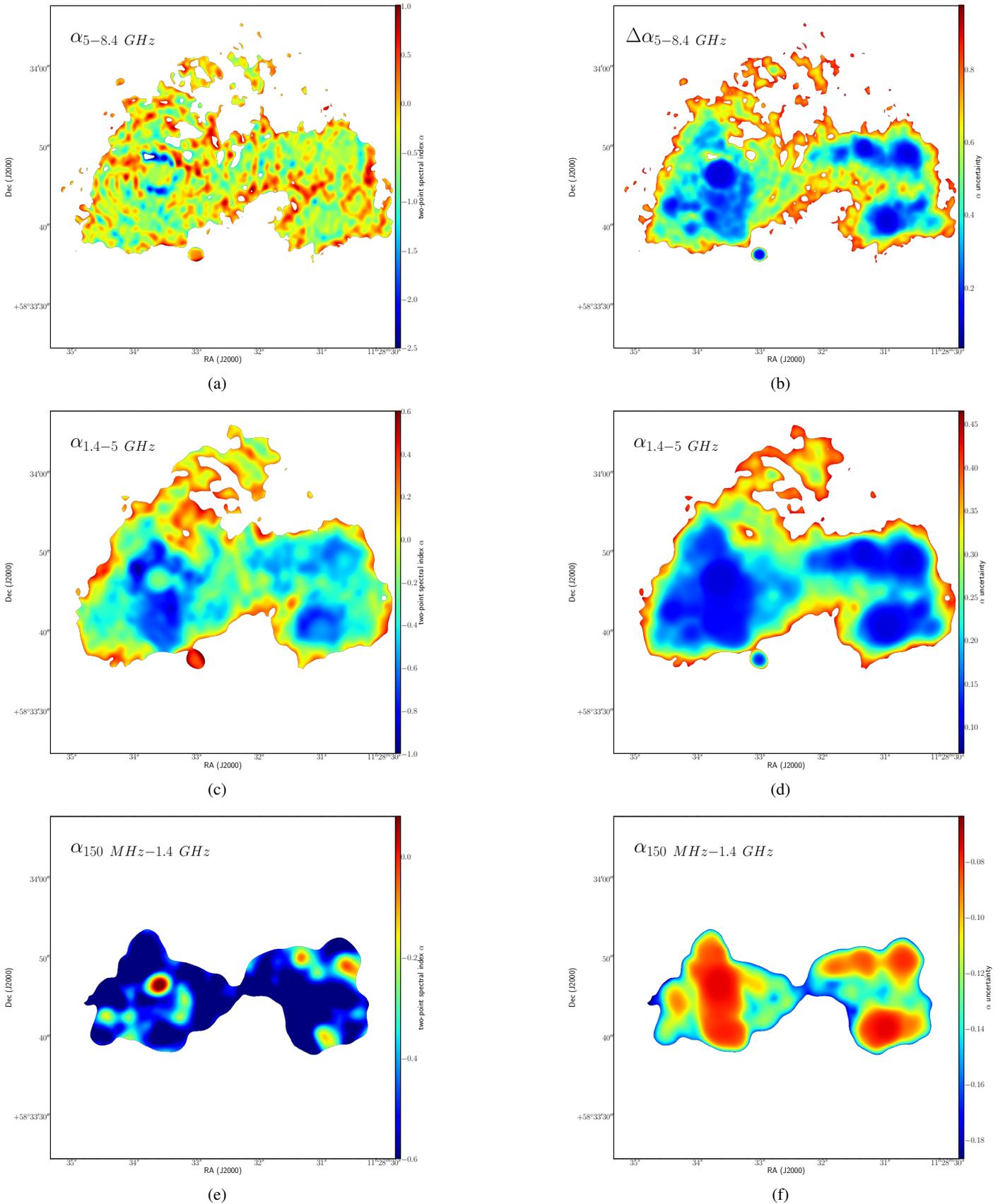

\begin{subfigure}{.43\textwidth}
  \centering
  \includegraphics[width=1.\linewidth]{spix_X-C.pdf}
  \caption{}
  \label{fig:spix_jvla_c-x}
\end{subfigure}%
\hspace*{\fill}%
\begin{subfigure}{.43\textwidth}
  \centering
  \includegraphics[width=1.\linewidth]{erorr_spix_X-C.pdf}
  \caption{}
  \label{fig:error_spix_jvla_c-x}
\end{subfigure} \\
\begin{subfigure}{.43\textwidth}
  \centering
  \includegraphics[width=1.\linewidth]{images/spix_C_L.pdf}
  \caption{}
  \label{fig:spix_jvla_l-c}
\end{subfigure}%
\hspace*{\fill}%
\begin{subfigure}{.43\textwidth}
  \centering
  \includegraphics[width=1.\linewidth]{erorr_spix_C-L.pdf}
  \caption{}
  \label{fig:error_spix_jvla_l-c}
\end{subfigure} \\
\begin{subfigure}{.43\textwidth}
  \centering
  \includegraphics[width=1.\linewidth]{spix_L-LOFAR.pdf}
  \caption{}
  \label{fig:spix_jvla_lofar-l}
\end{subfigure}%
\hspace*{\fill}%
\begin{subfigure}{.43\textwidth}
  \centering
  \includegraphics[width=1.\linewidth]{images/error_spix_L-LOFAR.pdf}
  \caption{}
  \label{fig:error_spix_jvla_lofar-l}
\end{subfigure} \\
\caption{Blanked spectral index maps (left) and its spectral uncertainties (right). Plots for the spectral index maps; (a) corresponds to  $\alpha_{(8.4GHz-5GHz)}$ at a resolution of $0.8"\times 0.8"$, (c) to $\alpha_{(5.0GHz-1.4GHz)}$ at a resolution of $1.4"\times 1.4"$, and (e) to $\alpha_{(1.4GHz-150MHz)}$ at a resolution of $1.4"\times 1.4"$ \textit{Right: } Total spectral index error (taking into account  Eq. \ref{eq:spectral_index_error}) at the frequencies given by the left image of each one. For more details, see Sect. \ref{sec:rayleigh-jeans}.}
\label{fig:spix}
\end{figure*}

We obtained two-point spectral index images ($\alpha$, according to the power law S$_{\nu}$=$\nu^{\alpha}$)  by combining our LOFAR and JVLA observations. We first used the previously convolved and regridded images at each frequency, as described in the previous subsection, and then applied the CASA task \texttt{immath} to the following pairs of images: 
150 MHz--1.4 GHz; 1.4 GHz--5.0 GHz; 5.0 GHz--8.4 GHz 
(left panels in Fig. \ref{fig:spix}).

We computed the pixel-wise spectral index uncertainty as in  \citet{kimtrippe,varenius2016}:

\begin{equation}
E(\alpha _{\nu1,2})=\frac{1}{\log(\nu_2/\nu_1)}\times \left[\frac{\sigma_{\nu_1}^2}{S_{\nu_1}^2} + \frac{\sigma_{\nu_2}^2}{S_{\nu_2}^2}\right]^{1/2}
\label{eq:spectral_index_error}
,\end{equation}

where $\sigma_{\nu_i}$ is the uncertainty in the total intensity at $\nu_i$. We blanked out all pixels in the spectral index maps that had an uncertainty larger than 3$\sigma$ in both frequencies, in the convolved and regridded images (see blanked out images in Fig. \ref{fig:spix}).

The spectral index maps in Fig. \ref{fig:spix} help us understand the nature of the emission of the different  regions. In the spectral index map between 5 and 8.4~GHz (Fig. \ref{fig:spix_jvla_c-x}) the general trend is within the range $\alpha_{8.4GHz-5GHz} \in (-0.3,-2.)$, suggesting that the whole galaxy is dominated by non-thermal synchrotron emission. The spectral index of the radio emission in the nuclear regions in Arp~299 starts to depart from the spectral index of the diffuse emission 
at some frequency between 1.4 and 5~GHz (Fig. \ref{fig:spix_jvla_l-c}). This trend becomes more evident
in Fig. \ref{fig:spix_jvla_lofar-l}, where the contrast between the spectral index, $\alpha_{1.4GHz-150MHz}$, clearly distinguishes the diffuse emitting regions,  $\alpha_{1.4GHz-150MHz} \leq$ -0.5, from the nuclear emission,  $\alpha_{1.4GHz-150MHz} \sim$ 0.0.
This different low-frequency behaviour of the synchrotron emitting regions (the nuclei display flat spectra, while the diffuse emission outside of the nuclei shows a rather steep spectrum)\ can be easily explained if for example    free-free absorption becomes increasingly relevant at frequencies around and below 1.4 GHz within the nuclear regions. In this case, the radio spectra will flatten in those regions,  diverging from the pure synchrotron behaviour observed in the regions outside the nuclei.

\subsubsection{Spectral energy distributions of compact components }
\label{sec:rayleigh-jeans}

We show in Table \ref{tab:indirect_result}  the flux density per solid angle, $S_{\nu}/\Omega$,  for each compact component and frequency of Arp~299, where we used the sizes obtained as described in Sect.~\ref{sec:integrated_flux}. We plot these values against frequency in Fig. \ref{fig:SED_nuclei} to  build up broadband SEDs between 150 and 8400 MHz. The spectra of all four components are inconsistent with a pure, unabsorbed synchrotron power law,  but show instead signs of flattening. In the next section we fit these spectra using models with mixed synchrotron and free-free emission and absorption.

\begin{table*}[]
\centering
\begin{threeparttable}
\caption{Fitted spectral model parameters for each compact component and derived physical quantities.}
\label{tab:indirect_result}
\begin{tabular}{llllllllllllllll}
\hline
   & $R_{ang}$                  & \multicolumn{4}{c}{S$_{\nu}$/$\Omega$}                                      & \multicolumn{1}{c}{$N_{cl}$} & \multicolumn{1}{c}{$\alpha$} & \multicolumn{1}{c}{$f_{th}$} & \multicolumn{1}{c}{$\nu_{t}$} & EM/10$^5$        & B$_{eq}$                   & \multicolumn{4}{c}{$t_{sync}$}                                         \\ 
       & \multicolumn{1}{c}{arcsec} & \multicolumn{4}{c}{mJy/arcsec$^{2}$}                                                                                                                                                             &                       &                                    &                              & GHz                           & pc$\, $cm$^{-6}$ & \multicolumn{1}{c}{$\mu$G} & \multicolumn{4}{c}{10$^3$yr} \\
   
   & \multicolumn{1}{c}{(1)} &  \multicolumn{4}{c}{(2)} &
   \multicolumn{1}{c}{(3)}                                    & \multicolumn{1}{c}{(4)}                              & \multicolumn{1}{c}{(5)}                           & \multicolumn{1}{c}{(6)}  & \multicolumn{1}{c}{(7)}  &  \multicolumn{1}{c}{(8)}  & \multicolumn{4}{c}{(9)} 
   \\ \hline
   &                            & \begin{tabular}[c]{@{}l@{}}150\\ MHz\end{tabular} & \begin{tabular}[c]{@{}l@{}}1.4\\ GHz\end{tabular} & \begin{tabular}[c]{@{}l@{}}5.0\\ GHz\end{tabular} & \begin{tabular}[c]{@{}l@{}}8.4\\ GHz\end{tabular} &                       &                                    &                              &                               &                  &                            & \begin{tabular}[c]{@{}l@{}}150\\ MHz\end{tabular} & \begin{tabular}[c]{@{}l@{}}1.4\\ GHz\end{tabular} & \begin{tabular}[c]{@{}l@{}}5.0\\ GHz\end{tabular} & \begin{tabular}[c]{@{}l@{}}8.4\\ GHz\end{tabular} \\ \cline{3-6} \cline{13-16} 
A  & 1.52                       & 32.4                                              & 37.5                                              & 29.6                                              & 22.3                                              & -                     & -0.33                              & 0.007                        & 0.23                          & 1.4              & 246                        & 154                                               & 17                                                & 16                                                & 12                                                \\
   &                            &                                                   &                                                   &                                                   &                                                   &                       & (0.11)                             & (0.004)                      & (0.08)                        &                  &                            &                                                   &                                                   &                                                   &                                                   \\ \arrayrulecolor{Silver} \arrayrulecolor{Silver} \arrayrulecolor{Silver} \arrayrulecolor{Silver} \cline{7-16} 
   &                            &                                                   &                                                   &                                                   &                                                   & {\it 5.2}                   & {\it -0.6}                               & {\it 0.18}                         & {\it 1.16}                          & {\it 42}               & {\it 286}                        & {\it 114}                                               & {\it 12}                                                & {\it 13}                                                & {\it 10}                                                \\
   &                            &                                                   &                                                   &                                                   &                                                   & {\it (0.8)}                 & {\it (fixed)}                            & {\it (0.04)}                       & {\it (0.08)}                        &                  &                            &                                                   &                                                   &                                                   &                                                   \\ \arrayrulecolor{black} \arrayrulecolor{black}  \hline
B  & 1.58                       & 20.9                                              & 11.3                                              & 5.9                                               & 4.3                                               & -                     & -0.56                              & 0.014                        & 0.18                          & 0.8              & 186                        & 260                                               & 28                                                & 24                                                & 18                                                \\
   &                            &                                                   &                                                   &                                                   &                                                   &                       & (0.02)                             & (0.002)                      & (0.01)                        &                  &                            &                                                   &                                                   &                                                   &                                                   \\ \arrayrulecolor{Silver} \arrayrulecolor{Silver} \arrayrulecolor{Silver} \cline{7-16} 
   &                            &                                                   &                                                   &                                                   &                                                   & {\it 1.8}                   & {\it -0.6}                               & {\it 0.05}                         & {\it 0.33}                          & {\it 2.4}              & {\it 192}                        & {\it 253}                                               & {\it 27}                                                & {\it 24}                                                & {\it 18}                                                \\
   &                            &                                                   &                                                   &                                                   &                                                   & {\it (0.2)}                 & {\it (fixed)}                            & {\it (0.01)}                       & {\it (0.03)}                        &                  &                            &                                                   &                                                   &                                                   &                                                   \\ \arrayrulecolor{black} \arrayrulecolor{black} \arrayrulecolor{black} \hline
C  & 2.09                       & 2.8                                               & 1.95                                              & 1.16                                              & 0.88                                              & -                     & -0.59                              & 0.089                        & 0.23                          & 1.4              & 106                        & 790                                               & 87                                                & 55                                                & 43                                                \\
   &                            &                                                   &                                                   &                                                   &                                                   &                       & (0.02)                             & (0.008)                      & (0.04)                        &                  &                            &                                                   &                                                   &                                                   &                                                   \\ \arrayrulecolor{Silver} \arrayrulecolor{Silver} \cline{7-16} 
   &                            &                                                   &                                                   &                                                   &                                                   & {\it 5.6}                   & {\it -0.6}                               & {\it 0.18}                         & {\it 0.24}                          & {\it 1.5}              & {\it 102}                        & {\it 903}                                               & {\it 97}                                                & {\it 61}                                                & {\it 47}                                                \\
   &                            &                                                   &                                                   &                                                   &                                                   & {\it (7.6)}                 & {\it (fixed)}                            & {\it (0.06)}                       & {\it (0.03)}                        &                  &                            &                                                   &                                                   &                                                   &                                                   \\ \arrayrulecolor{black} \arrayrulecolor{black} \hline
C' & 1.96                       & 3.18                                              & 2.1                                               & 1.4                                               & 1.2                                               & -                     & -0.39                              & 0.050                        & 0.15                          & 0.6              & 102                        & 883                                               & 95                                                & 60                                                & 46                                                \\
   &                            &                                                   &                                                   &                                                   &                                                   &                       & (0.02)                             & (0.008)                      & (0.04)                        &                  &                            &                                                   &                                                   &                                                   &                                                   \\ \arrayrulecolor{Silver} \arrayrulecolor{Silver} \cline{7-16} 
   &                            &                                                   &                                                   &                                                   &                                                   & {\it 5.0}                   & {\it -0.32}                              & {\it 0.038}                        & {\it 0.122}                         & {\it 0.4}              & {\it 97}                         & {\it 903}                                               & {\it 97}                                                & {\it 61} & {\it 47}                                                \\
   &                            &                                                   &                                                   &                                                   &                                                   & {\it (fixed)}               &     {\it (0.07}                                & {\it (0.003)}                      & {\it (0.001)}                       &                  &                            &                                                   &                                                   &                                                   &                                                   \\ \arrayrulecolor{black} \hline
\end{tabular}
\begin{tablenotes}
\footnotesize
\item[Notes]
Columns 2 to 6 give the size and estimated surface brightness versus frequency of each component. Columns 7 to 10 give the spectral model fit parameters  split into separate rows for the continuous free-free model fits (normal typeface) and clumpy free-free model fits (italics). Values in brackets correspond to the estimated uncertainties of the values above. For the clumpy model some of theses values are held fixed (see \ref{sec:clumpy_model}). Finally, Cols. 11 to 15 give the derived physical parameters for each model. Each column is described in the table notes with the respective index number. [1] Angular radius of each component, measured as described in Sect. \ref{sec:integrated_flux}. [2] Flux density per solid angle measured for each component, calculated as  described in Sect. \ref{sec:integrated_flux}. [3] Mean number of clumps intercepted per LOS assuming the clump model. This is only given for the clumpy model rows (see table caption). [4] Spectral index of synchrotron emission,  as described in Sect. \ref{sec:free-free_ISM}. [5] Thermal emission as a fraction of synchrotron emission at 1~GHz, as described in Sect. \ref{sec:free-free_ISM}. [6] Frequency at which free-free opacity through source equals unity (see Sect. \ref{sec:free-free_ISM} and Appendix A). [7] Emission measure along the  LOS, estimates as described in Sect. \ref{sec:free-free_ISM}. [8] Equipartition magnetic field, estimated as  described in Sect. \ref{sec:magnetic_field}. [9] Synchrotron lifetime of electrons assuming an equipartition magnetic field estimated as explained in Sect. \ref{sec:energy_losses}.
\end{tablenotes}
\end{threeparttable}
\end{table*}

\subsection{Magnetic field}
\label{sec:magnetic_field}

We show in Table \ref{tab:indirect_result} the minimum equipartition magnetic field for the nuclei of Arp~299, which is  the field  obtained by assuming equipartition between the cosmic ray energy and the magnetic field energy \citep{pacholczyk},

\begin{equation}
B_{eq} = \left(4.5c_{12}/\phi\right)^{2/7}\left( 1 + \psi\right)^{2/7}R^{-6/7}L_R^{2/7}
\label{eq:magnetic_field}
,\end{equation}

where $c_{12}$ is a slowly varying function of the spectral index, $\alpha$; $\phi$ is the filling factor of fields and particles; $R$ is the linear radius of the emitting region; $L_R$ is the radio luminosity of the source; and $\psi$ is the ratio of the total heavy particle energy to the electron energy.
Depending on the mechanism that generates the relativistic electrons, this ratio  ranges between $\psi \approx$1 and $\psi \approx m_p/m_e \approx$2000 (where $m_p$ and $m_e$ are the proton and electron mass, respectively). For simplicity, we used here $\phi$ = 0.5 and $\psi$ = 100.

We calculated the magnetic field for each nuclear region by following the prescription of Eq. \ref{eq:magnetic_field}.  
 As the physical size, $R$, of each component, we used the values in Table \ref{tab:indirect_result}. We note that these values correspond to the continuous case discussed in Sect. \ref{sec:free-free_ISM} since in the clumpy model (Sect. \ref{sec:clumpy_model}) the size of each clump is not considered.

\section{Discussion}

\subsection{Emission models for compact components}
\label{subsec:emission_models}
Radio emission in LIRGs comes from  a combination of synchrotron and free-free emission. Synchrotron emission arises 
from relativistic electrons accelerated in supernovae (with possible contributions from buried AGN and associated radio jets), interacting with magnetic fields. The free-free emission component is bremsstrahlung emission from ISM gas ionized by massive stars. While the free-free thermal gas  makes only a small emission contribution at gigahertz frequencies and below, it can have a large effect on the overall spectrum at low radio frequencies because it can effectively absorb background synchrotron emission. Such low-frequency absorption effects of thermal ionized gas can readily explain the peaked spectra observed in the radio spectra of the compact components in Arp~299 (Fig. \ref{fig:SED_nuclei}).

The standard model of radio emission in LIRGs \citep{condon91,Condon92} assumes that both the synchrotron and free-free media are smooth and co-extensive. 
In Section \ref{sec:free-free_ISM} we apply this model to fit the observed spectra of the Arp~299 compact components and discuss the results. In reality, it is probable that the free-free emission and/or absorption in LIRGs is not smooth, but is instead concentrated within discrete HII regions \citep{lacki2014}. Both \citet{lacki2014} and \citet{conway} have presented spectral models which take into account this fundamentally clumpy nature of the free-free absorbing medium. In Sect. \ref{sec:clumpy_model} we fit models based on the clumpy medium formalism of \citet{conway} and compare these results to those obtained by fitting the continuous model.

\subsubsection{Continuous free-free medium model fitting}
\label{sec:free-free_ISM}
The continuous free-free  medium model \citep{condon91}, following the formulation of  \citet{varenius2016}, predicts a radio brightness versus frequency given by
\begin{equation}
     S_{\nu,\Omega} = 7.22\,\nu^2\, (T_e/10^{4} {\rm K})\,(1-e^{-\tau_{\nu}})\,\left[1 + {f_{th}}^{-1} \nu^{\alpha + 0.1}\right]
\label{eq:condon91}
,\end{equation}
\noindent where $S_{\nu,\Omega}$ is the surface brightness in units of  mJy\,arcsec$^{-2}$, the frequency $\nu$ in units of GHz, and 
$T_e$ is the free electron temperature in Kelvin; $f_{th}$
is the ratio of thermal to synchrotron emission at 1~GHz 
(assuming both emission mechanisms are optically thin, i.e. ignoring absorption effects); and $\alpha$ is the non-thermal synchrotron spectral index. 
The free-free opacity versus frequency in this model follows 
$\tau_{\nu}=\left(\nu/\nu_t\right)^{-2.1}$,
where
$\nu_{t}$ 
is the turnover frequency at which free-free opacity through the  continuous source medium is unity. If we fix $T_e=10^{4}$ K the model has three free parameters ($\alpha$, $f_{th}$, $\nu_{t}$);  the turnover frequency   
\citep{Metzger1967} is given by

\begin{equation}
\nu_{t} ({\rm GHz}) = \left[ \frac{EM ({\rm pc\, cm^{-6}})}{3.05\times 10^{6}}\right] ^{1/2.1}
\noindent 
 \label{equ:turnEM}
,\end{equation}

\noindent where the 
emission measure (EM) through the source medium equals the integral of  $n_{e}^{2}$ through  the source where $n_{e}$ is the thermal electron density.

\begin{figure*}
\centering
\begin{subfigure}{0.5\textwidth}
  \centering
  \includegraphics[width=1.\linewidth]{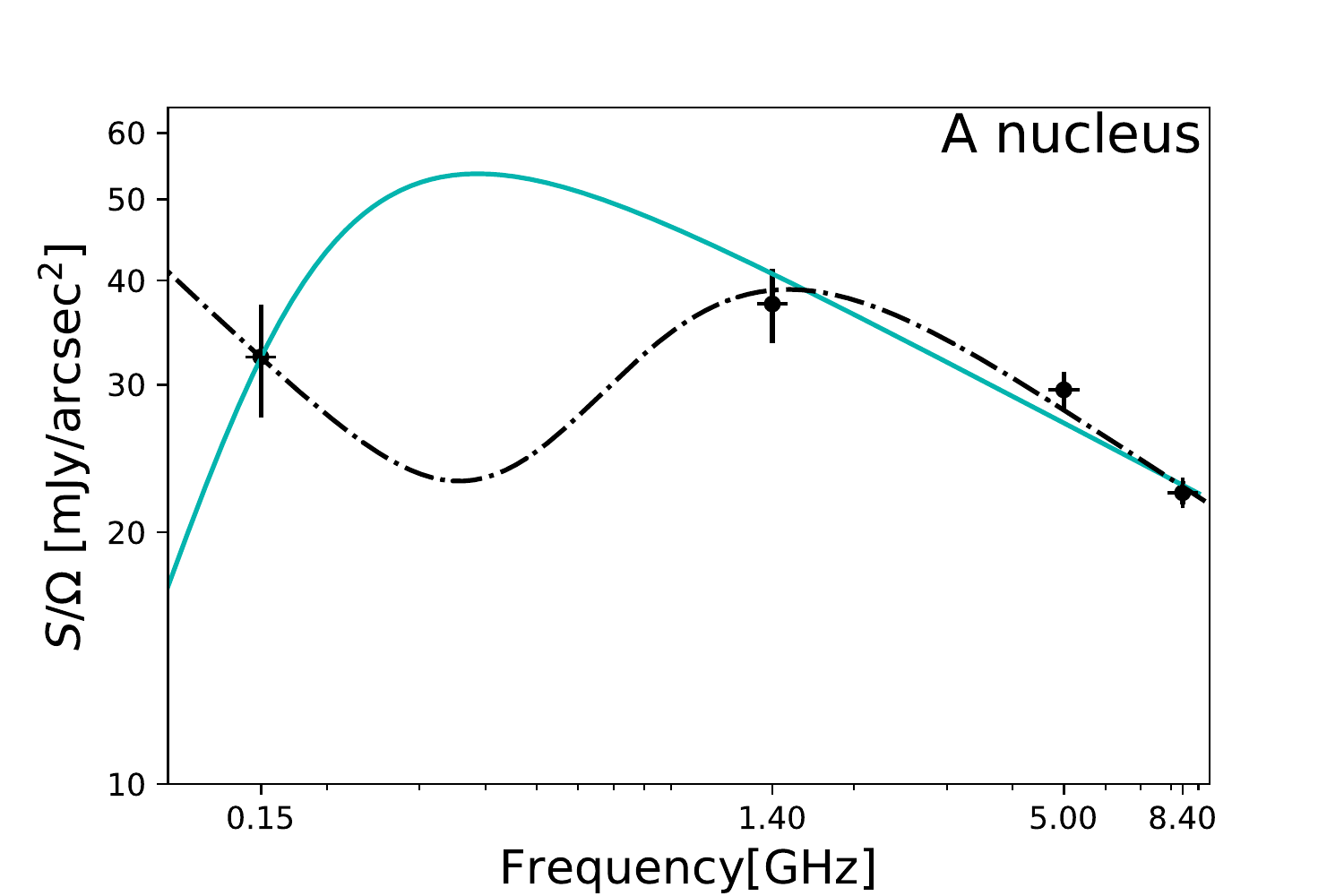}
  \caption{A nucleus}
  \label{fig:SED_A}
\end{subfigure}%
\begin{subfigure}{0.5\textwidth}
  \centering
  \includegraphics[width=1.\linewidth]{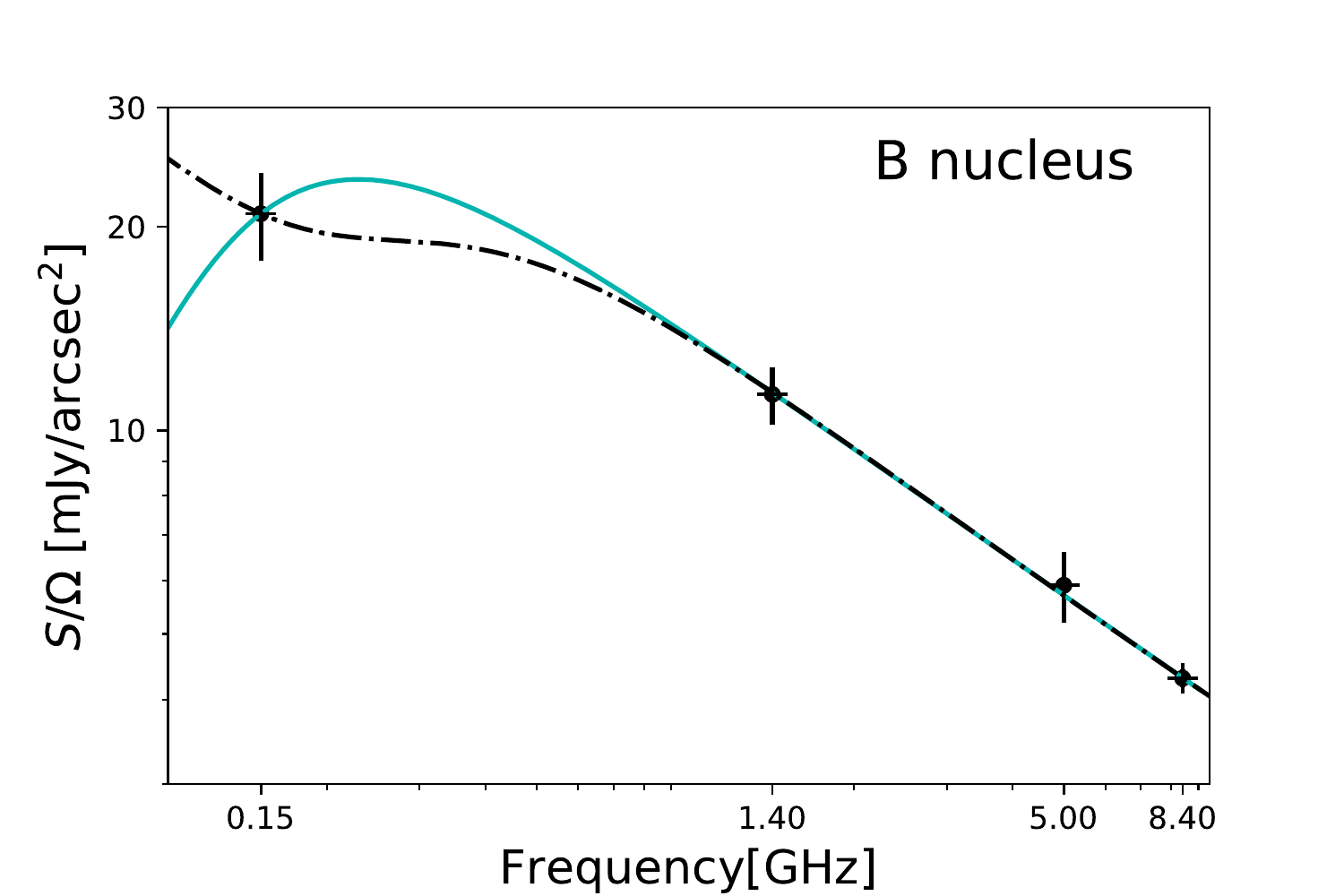}
  \caption{B nucleus}
  \label{fig:SED_B}
\end{subfigure}\\
\begin{subfigure}{0.5\textwidth}
  \centering
  \includegraphics[width=1.\linewidth]{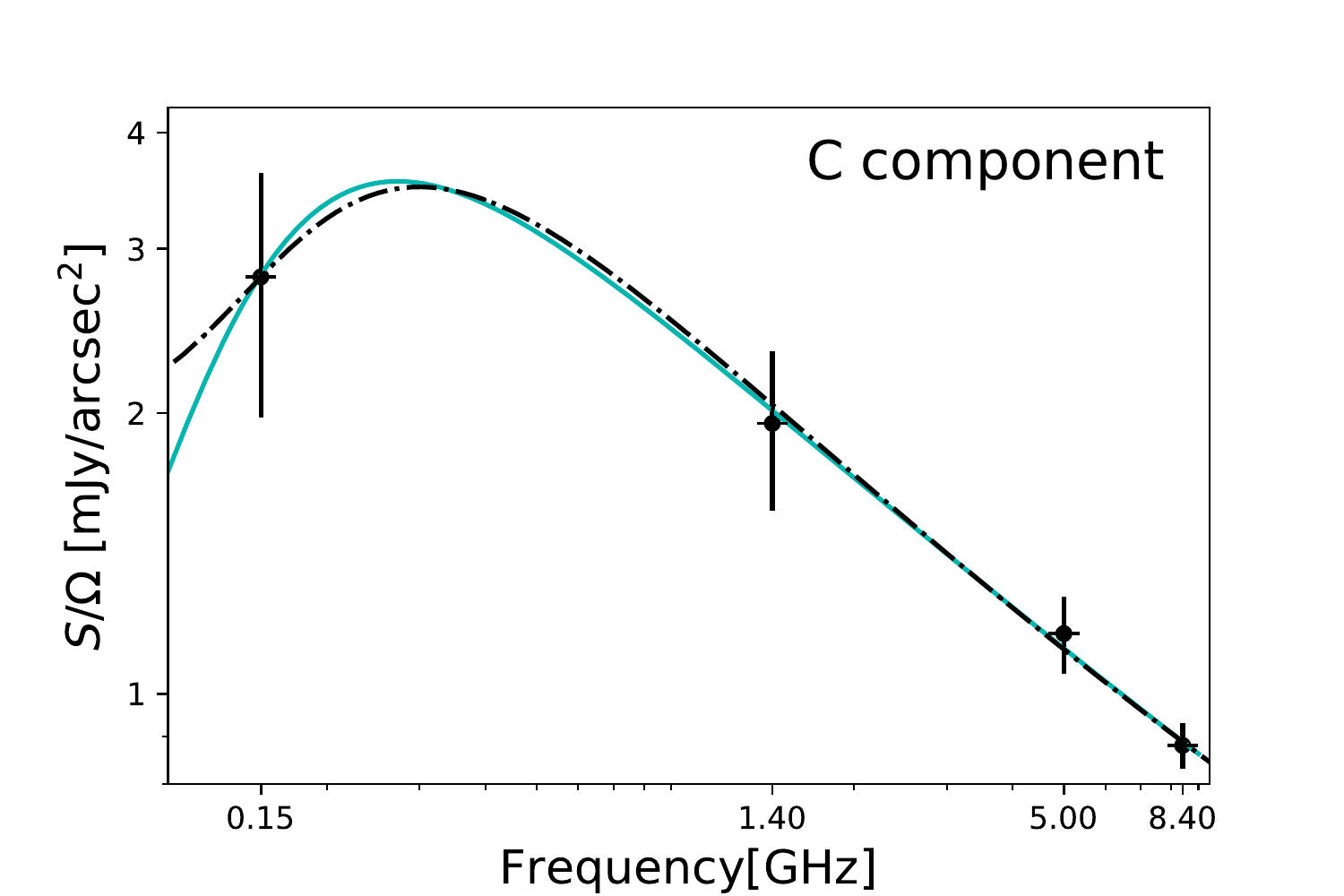}
 \caption{C nucleus}
  \label{fig:SED_C}
\end{subfigure}%
\begin{subfigure}{0.5\textwidth}
  \centering
  \includegraphics[width=1.\linewidth]{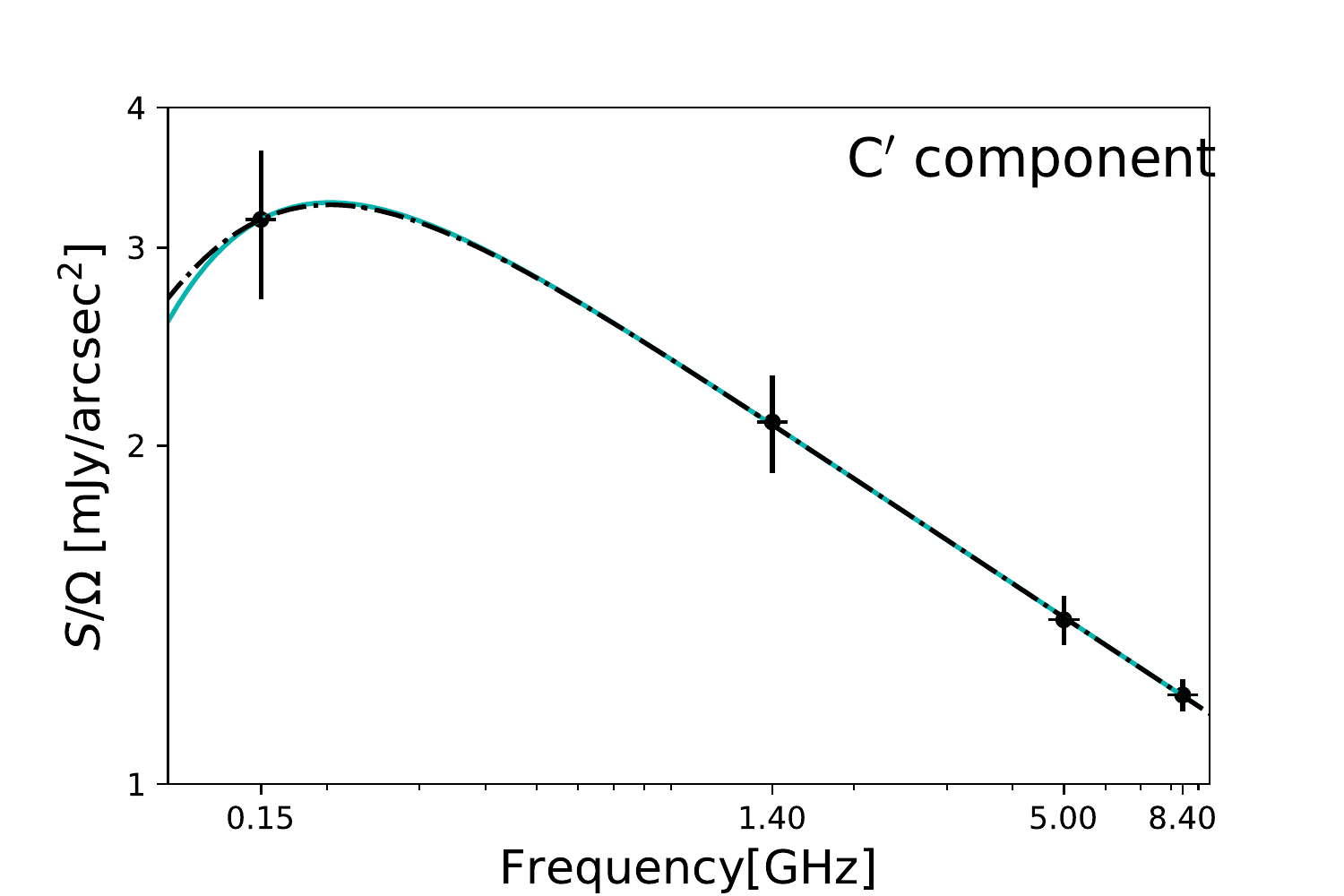}
  \caption{C' nucleus}
  \label{fig:SED_Cprime}
\end{subfigure}\\
\caption{SEDs of the nuclei in Arp~299. Their sizes are  defined by  50$\%$ of the peak in the convolved (1.4 arcsec$^2$ beam) of 1.4 GHz JVLA image. The solid blue lines and dash-dotted lines correspond to the best fit to the smooth continuum and clumpy models, respectively.}
\label{fig:SED_nuclei}
\end{figure*}

 We used the model given in  Eq. \ref{eq:condon91} to fit the spectrum  of each compact component, setting the electron temperature $T_{e}=10^{4}$ K and solving for the three remaining free parameters. 
 We used the Python fitting \texttt{lmfit} package to  minimize the square of the difference between the data and the different models. We show the results of this fit in Fig. \ref{fig:SED_nuclei} and Table \ref{tab:indirect_result}.  Generally, the  continuous model provides excellent fits to the data with the possible exception of component A. For this component we find that in order to accommodate both the 150 MHz and the higher frequency spectral points a free-free turnover is required at the relatively low frequency of 230 MHz; with the consequence that at 1.4 GHz and above the model spectrum is very close to power law. The resulting straight high-frequency spectrum  cannot therefore provide a  good fit to the apparent spectral curvature implied by  the 1.4, 5, and 8 GHz spectral points.
 
We also find that our  model fitting requires  synchrotron spectral indices of $-0.33\pm 0.11$ and $-0.39\pm0.02$
 for components A and C' respectively, which  are  significantly flatter spectral indices than are normally observed in star-forming galaxies ($\alpha \approx -0.8$; e.g. \citealt{HerreroIllana2017,PerezTorres2021}). The flatter spectral indices required for components A and C' in Arp~299 could be explained (see Sec. \ref{sec:energy_losses}) by the effects of bremsstrahlung and/or ionization loss effects  on the energy spectrum of the synchrotron emitting relativistic electrons in dense regions. Such loss mechanisms can flatten the energy spectrum of relativistic electrons at lower  energies, in turn giving  flattened synchrotron spectra below a critical frequency. Given the relatively high densities implied by the EM values obtained towards the four compact regions, a flattened spectral index  around 1.4 GHz and below is plausible (see Sect. \ref{sec:energy_losses}). More exact modelling, which is beyond the scope of this paper, would take into account the combined effects of  bremsstrahlung, ionization, and synchrotron losses, which in general are expected to generate a curved synchrotron spectrum rather than one that is a single  power law at all frequencies, as assumed by Eq. \ref{eq:condon91}. 
Another potential reason for the relatively flat radio spectral index of the A nucleus could be the existence of an AGN  (\citealt{perez-torres2010}). This AGN is  very faint, however,  about 1 mJy at centimetre wavelengths, and it therefore makes a negligible contribution to the overall radio spectral index of the A component.

Fitting  Eq. \ref{eq:condon91} to the four compact components shows that all require similar free-free absorption frequencies  close  to 0.2 GHz. This implies  for each component an EM of $\sim\,10^5$ pc~cm$^{-6}$ (see Eq. \ref{equ:turnEM}). This value is more than  seven orders of magnitude larger than the EM of the Milky Way (10$^{-2}$ pc~cm$^{-6}$;  \citealt{nakashima}), consistent with the expectation  of a much denser ISM in Arp~299. Somewhat surprising, however, is the 
similarity of the fitted turnover frequencies of the four components given that they span an order of magnitude in radio surface brightness. The empirically established correlation between radio synchrotron brightness and star formation rate (SFR) per unit area, combined with the expectation that the free-free EM also scales with area SFR \citep{Murphy2011}, means that we would expect brighter radio components to also have larger EM values, and hence higher turnover frequencies. Another way to view the same issue is that the fitted 1 GHz  thermal fractions  $f_{th}$ for the bright sources A and B are only 0.7\%  and 2\%, respectively, significantly smaller\footnote{The thermal fraction of component A is comparable to that estimated by \citet{varenius2016} for the high radio surface brightness East and West nuclei of Arp~220 (thermal fractions 0.8\% and 0.4\%, respectively)}  than for C and C'  or for normal star-forming galaxies ($\sim$10\%;  \citealt{Condon92}, \citealt{Niklas97}). Such small thermal fractions for A and B compared to C and C' are forced by the model fitting   because these bright components have high synchrotron brightness at 1 GHz, and yet their thermal gas content must be the same as for the weaker
components C and C' because all components have similar turnover frequencies.
The required anomalously low $f_{th}$ in the bright components A and B could theoretically arise if for some reason these components  have  an overabundance of synchrotron emission relative  to their SFRs. However, this explanation is   in conflict with A and B being much brighter in the IR  (see \citealt{PerezTorres2021}), implying that these components have significantly larger SFR. A more plausible explanation for the small $f_{th}$  fractions in A and B is that the synchrotron emission closely follows SFRs, but that the  free-free    EM is partially suppressed in these bright components. A possible reason  could be dust absorption, as suggested by 
\citet{BarcosMunoz15} to explain a similar suppression of free-free emission in Arp~220. Dust absorption in dense regions could reduce the number of high-energy photons available to ionize gas, which in turn reduces the EM and the turnover frequency.

In summary, we find that the continuous free-free absorption model provides excellent fits to the spectral data for the four compact components. However, for the bright components A and B these fits require a mechanism (possibly dust absorption of ionizing photons) to suppress the amount of free-free emitting/absorbing gas. Additionally, in two components (the brightest component A and  component  C')  relatively flat synchrotron spectral indices (i.e. $-0.33$ and $-0.39$, respectively) are required. Such spectral indices could  be explained, however,  by the effects of bremsstrahlung and ionization losses in these dense regions. 

\subsubsection{Clumpy HII region free-free medium model fits}
\label{sec:clumpy_model}

Lacki (2013) has argued that in (U)LIRGs free-free gas is confined within discrete compact HII regions  surrounding high-mass stars or super star clusters. Such a physical origin for thermal gas contrasts with the assumption of the continuous free-free absorption model fitted in Sect. \ref{sec:free-free_ISM}. Nevertheless it can be shown (see \citealt{conway} and Appendix A) that the continuous model provides a good approximation to the emerging spectrum as long as individual clumps remain optically thin down to the lowest frequency observed.  

In this section we present fits to the observed compact component spectral data using a model that explicitly takes into account the effects of clumping and which therefore applies even when individual clumps are optically thick. Appendix A uses the formalism of \citet{conway}  to derive an expression for the clumpy spectrum (Eq. \ref{eq:clumpform}). For fixed electron temperature this expression has four free parameters, three of which are shared with the continuous case, namely the thermal/synchrotron ratio at 1 GHz
($f_{th}$), the synchrotron spectral index 
($\alpha$),  and the frequency ($\nu_{t}$)  at which total free-free opacity through the clump population is unity (roughly the turnover frequency). The fourth  free parameter is  $N_{cl}$
the mean number of clumps intercepted per line of sight (LOS). As described in Appendix \ref{sec:app} down to the frequency at which individual clumps become optically thick (i.e. $\nu_{turn-up} = \nu_{t}N_{cl}^{(-1/2.1)}$) the clumpy medium spectrum closely tracks that of a continuous medium spectrum, which has  the same values as $f_{th}$, $\alpha$, and $\nu_{t}$. At the frequency $\nu_{turn-up}$, however,  the emission  reaches a local minimum, below which it increases again, eventually becoming power law  with spectral index  $\alpha$
(see  an example of  the spectrum for component A in Fig. \ref{fig:SED_nuclei}).

 We fitted  the observed spectra of the compact components in Arp~299 using the clump spectral model of Eq. \ref{eq:clumpform}. The four free  parameters of the clump model equal the number of  spectral points to fit, so that if all parameters are adjusted the  fitting has no degrees of freedom. In order to stabilize the fitting and not to be too sensitive to noise, we therefore chose to fix one of the parameters and fit for the other three. For components A, B, and C we fixed the spectral index at  $\alpha=-0.6$ (acceptable fits were found for all $\alpha >-0.5),$  which is in the range of synchrotron spectral index typically observed in star-forming galaxies. For component C' it was not possible to get a good fit assuming this spectral index, so instead the number of clumps per LOS was fixed at $N_{cl}=5$ forcing  a solution similar to the continuous model.  
 
 The best fitting clump models are shown by the dash-dotted lines in Fig. \ref{fig:SED_nuclei}, which use the  model parameters shown in italics in Table \ref{tab:indirect_result}. The clump fits shown for components A and B are quite different from the continuous models for these components; they show a significantly larger initial turnover frequency, followed at lower frequencies by a spectral `turn-up' that occurs above the frequency of the LOFAR data point. In contrast, the clump fits for C and C' are   very similar to the continuous model over the range of the observations. For these components,  which have relatively low turn-over frequencies, the  clump induced spectral turn-ups occur  well below the lowest observed data point. It is interesting to note in Fig. \ref{fig:SED_nuclei} how the turnover frequencies for the clump models show a decrease as we progress from brighter to weaker components. This is as expected assuming an  approximately constant ratio of  free-free to synchrotron, which implies that brighter components have more free-free gas and hence have higher turnover frequencies.
  For components A and B it should be noted that 
there  are alternative clump models that provide  fits to the data that are just as  good as those plotted in Fig. \ref{fig:SED_nuclei}. These alternative fits, like those for C and C', track the continuous model closely over the range of frequencies observed and have turn-up frequencies significantly below the lowest observed data point. Since these alternative clump models do not show the expected progression of turnover frequency with peak component brightness and, in any case, these have essentially the same physical properties  of the continuous model (e.g. EM), we do not discuss these alternative clump models further. 
 
If we compare in detail the clump fit shown in Fig. \ref{fig:SED_nuclei} for component A (blue curve) with the continuous model fit from subsection \ref{sec:free-free_ISM} (dashed black curves) we find that  (1)  the clump fits gives a better, though still not perfect, fit  to the apparent spectral curvature defined by the three highest frequency points; (2)  for the clump case a good solution exists having a synchrotron spectral index ($\alpha = -0.6$) closer to values typically observed in intense star-forming galaxies \citep[i.e.][]{basu2} than the flat value of -0.33 required for the continuous model; and  (3) the required thermal fraction for the clump model is 18\% compared to the 0.7\% for the continuous model. This higher thermal fraction is much closer to the typical value of about 10\% found in normal  star-forming galaxies. Turning to component B we find that (1) the clump and continuous fits are equally good; (2) the synchrotron spectral indices are very similar for the two  models, with both being close to values commonly observed in star-forming galaxies; (3) the required thermal fraction is 5\% for the clump model rather than the 1.8\% required for the continuous model, which again is closer for the clump case to the value commonly observed in star-forming galaxies. Overall, these   results  show that the clump model fits the spectral data either somewhat better than (component A) or equally as well as (components B, C, and C')  the continuous model. The thermal fractions derived from the clump model are more similar between components and closer to values typical of star-forming galaxies. Finally, good clump fits can be made assuming (for components A, B, and C) a synchrotron spectral  index of -0.6 which is more consistent with the normal range observed in star-forming galaxies (C' still requires a flat spectral index even in the clump case).  While the above arguments broadly support the clump model  we note that alternative explanations exist  (see Sect. \ref{sec:free-free_ISM}) within the frame of  the continuous model for their somewhat non-standard spectral indices  and thermal fractions.

\subsubsection{Deciding between continuous and clumpy free-free models}
\label{sec:decide_model}

 It is interesting to consider possible future observations that could clearly distinguish between the continuous and clump models for A and B. One systematic difference between the two models is in  their thermal fractions (for component A   0.7\% and 18\% for continuous and clump models, respectively). This in turn  predicts different amounts of flat spectrum high-frequency free-free emission in the two cases, which could be distinguished via 
high-frequency VLA observations. Another clear difference in spectral shape for the two models occurs between 150 MHz and 1.4 GHz, and this could potentially  be distinguished using uGMRT observations at 200, 600, and 800 MHz. Additionally, the clump and continuous models  predict respectively negative and positive spectral indices at 150 MHz. A spectral index  measured internally over the LOFAR HBA observing band should in principle be able to distinguish between these different  predictions, but   the internal data calibration of the present LOFAR data is not of high enough accuracy to do this.  Finally, there is the possibility of using future LOFAR Low Band Array international baseline observations  to observe around 60 MHz. There is a clear prediction from the clump model that at this frequency emission from components A and B should be significantly brighter than at 150 MHz, while  the continuous model predicts the opposite (Fig. \ref{fig:SED_nuclei}).

\subsection{Energy losses and the spectral index of the Arp~299 nuclei}
\label{sec:energy_losses}

If relativistic electrons are injected with a power law $p=2.1$
($N(E) \propto E^{-p}$) and the cooling of electrons is due to synchrotron and/or inverse Compton (IC)
losses, then the index $p$ would steepen by unity (i.e. $\cal P$ = $p+1$). In the absence of other energy losses, the observed radio spectrum $S_\nu \propto \nu^{\alpha}$ steepens, and becomes equal to
$\alpha = (1 - \cal P)$/2 = $- p/2$.  If $p=2.1$, then $\alpha = -1.05$. However, this is
not observed, as indicated in Table \ref{tab:indirect_result}, which shows that all nuclei have radio spectra
significantly flatter than $\alpha = -1.0$.  Hence, the observed spectral flattening
suggests that electron losses other than radiative losses (synchrotron and IC) are
likely affecting the observed spectrum.

To determine the importance of synchrotron cooling relative to other losses, we can compare the lifetimes of electrons.  The synchrotron cooling timescale for cosmic ray (CR) electrons emitting at frequency $\nu$ is

\begin{equation}
t_{\rm synch}\approx 1.4\times10^5\,B_{100}^{-3/2}\nu^{-1/2}\,\,{\rm yr},
\label{eqn:tSynch}
\end{equation}

where $B_{100}=B/100$~$\mu$G is the minimum equipartition magnetic field, \Beq\, for
each of the nuclei in Arp~299,  
and $\nu$ is the critical frequency at which we are calculating 
the lifetime of the electrons, in GHz.

The lifetimes of electrons for the Arp~299 nuclei are in accordance
with other studies of LIRGs at similar frequencies (see e.g.
\citealt{romero-canizales1}), with the most long-lived ones in the B nucleus at the lower
frequencies of LOFAR. 

The synchrotron lifetime of the electrons must be compared against  escape, ionization, and bremsstrahlung losses.
For Arp~299-A, we know there is a starburst-driven outflow with a speed $v_w \sim (370 -
890)$ km\,s$^{-1}$ \citep{ramirez-olivencia}. Winds, whether AGN or
starburst-driven, can advect cosmic rays out of their host regions on relatively short
timescales. The corresponding lifetime is  

\begin{equation}
t_{\rm wind} = h/v_w \approx 2.4\times10^5\, h_{100}\, v_{w,400}^{-1}\, {\rm yr},
\label{eqn:twind}
\end{equation}

where $h = 100\,h_{100}$ pc is the scale height of the nuclear disk, which is about 100 pc for compact LIRGs \citep[e.g.][]{lacki2010a,herrero-illana}, and $v_w = 400\,v_{w,400}$km\,s$^{-1}$.
For the magnetic field values expected to exist in dense starbursts like Arp~299-A ($\gtrsim$100 $\mu$G and larger), $t_{\rm synch}$ is shorter than even the advection timescale (Eq. \ref{eqn:twind}), and therefore escape losses can be neglected, compared to synchrotron losses.

The electron lifetime due to inverse Compton losses can be written as \citep[e.g.][]{lacki2010a}

\begin{equation}
\label{eqn:tIC}
t_{\rm IC}\approx 5.7\times10^8\,B_{100}^{1/2}\nu_{\rm GHz}^{-1/2}U_{\rm ph,\,-12}^{-1}\,{\rm yr}
,\end{equation}
where $U_{\rm ph} = 10^{-12}\,U_{\rm ph,\,-12} $\,ergs cm$^{-3}$ is the photon energy density.

\begin{figure}[h!]
\includegraphics[width=1.\linewidth]{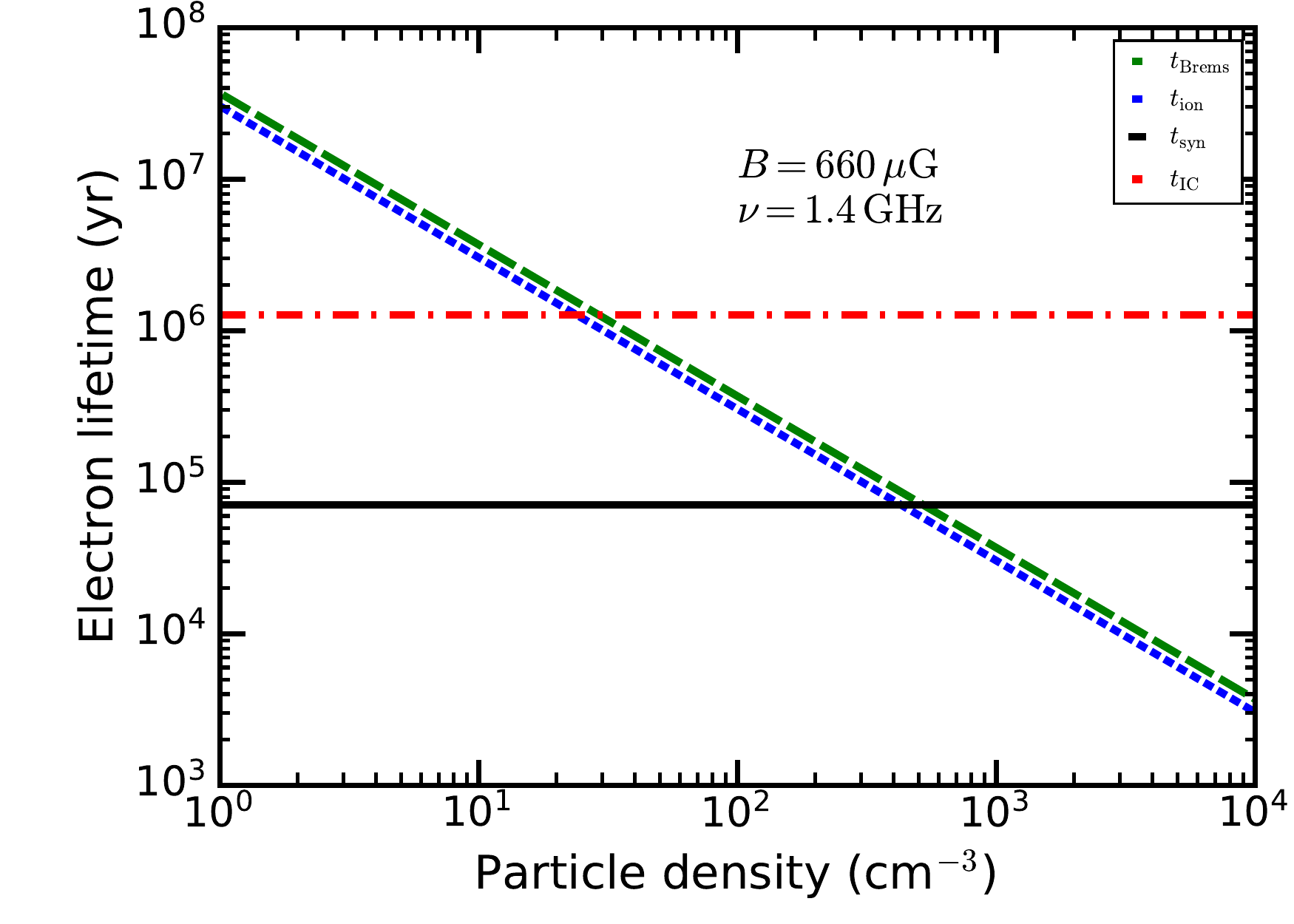}
\caption{Lifetime of electrons (in yr) emitting at 1.4 GHz, as a function of the particle number density (in cm$^{-3}$), and subjected to bremsstrahlung (long dashes), ionization (short dashes), synchrotron (solid line), and inverse  Compton (dash-dotted line) losses.  
Synchrotron and inverse Compton losses are appropriate for the A nucleus. 
We note that at low particle densities, which are appropriate for most extra-nuclear regions in galaxies,  radiative losses (i.e. synchrotron and IC) dominate, while at high particle densities like those encountered in the nuclear regions of LIRGs, bremsstrahlung and ionization losses dominate (see main text for details). }
\label{fig:electron_lifetimes}
\end{figure}

In LIRGs synchrotron cooling is expected to dominate over IC cooling. However, we
cannot neglect bremsstrahlung and ionization losses of CR electrons, whose 
efficiency increases with  the density of the ISM. 
Specifically, the bremsstrahlung and ionization energy loss timescales are  

\begin{equation}
\label{eqn:tBrems}
t_{\rm brems}\approx 3.7\times10^7\,n_e^{-1}\,{\rm yr},
\end{equation}

and
\begin{equation}
\label{eqn:tIon}
t_{\rm ion}\approx 6.6 \times 10^7\,B_{100}^{-1/2}\nu_{\rm GHz}^{1/2}
n_e^{-1} \,\,{\rm yr}.
\end{equation}

Bremsstrahlung and ionization  processes can flatten significantly  the equilibrium electron 
spectra. 
If bremsstrahlung losses dominate over all other losses, the injected spectrum is not modified (i.e. ${\cal P} = p $ and $\alpha =  (1 - p)/2$; \citealt{pacholczyk}). For $p = 2.1$, the spectral index would then become $\alpha = -0.55$.
Similarly, 
if ionization losses are the only relevant ones,  ${\cal P} = p - 1$ and $\alpha = 1 - p/2$ (i.e. $\alpha = -0.05$ for $p = 2.1$; \citealt{pacholczyk}).
When $t_{\rm synch} = t_{\rm
brems}$ at some energy, and all other losses are negligible, 
then ${\cal P} = p + 1/2$
and $\alpha = 1/4 - p/2 $ \citep{lacki2010a}.  Similarly, when $t_{\rm synch} = t_{\rm ion}$ and there are
no other losses ${\cal P} = p$ and $\alpha = (1- p)/2$ \citep{lacki2010a}.
For $p = 2.1$ we have $\alpha = -0.8$ and $\alpha = -0.55$ when  bremsstrahlung and ionization losses, respectively, are as important as synchrotron losses.

As an illustrative example, we show in  Fig. \ref{fig:electron_lifetimes} the different timescale losses for the A nucleus. For particle densities below $\sim$500 cm$^{-3}$, synchrotron losses dominate the emission at 1.4 GHz, and therefore the observed radio spectrum becomes steep. However, for particle densities above $\sim 500$
cm$^{-3}$, as seems to be the case in Arp~299-A and, most likely, for all other nuclei, bremsstrahlung and ionization losses can be as important, or even dominate over synchrotron losses, so that the radio spectrum flattens significantly.

\begin{figure*}
    \centering
    \includegraphics[width=\textwidth]{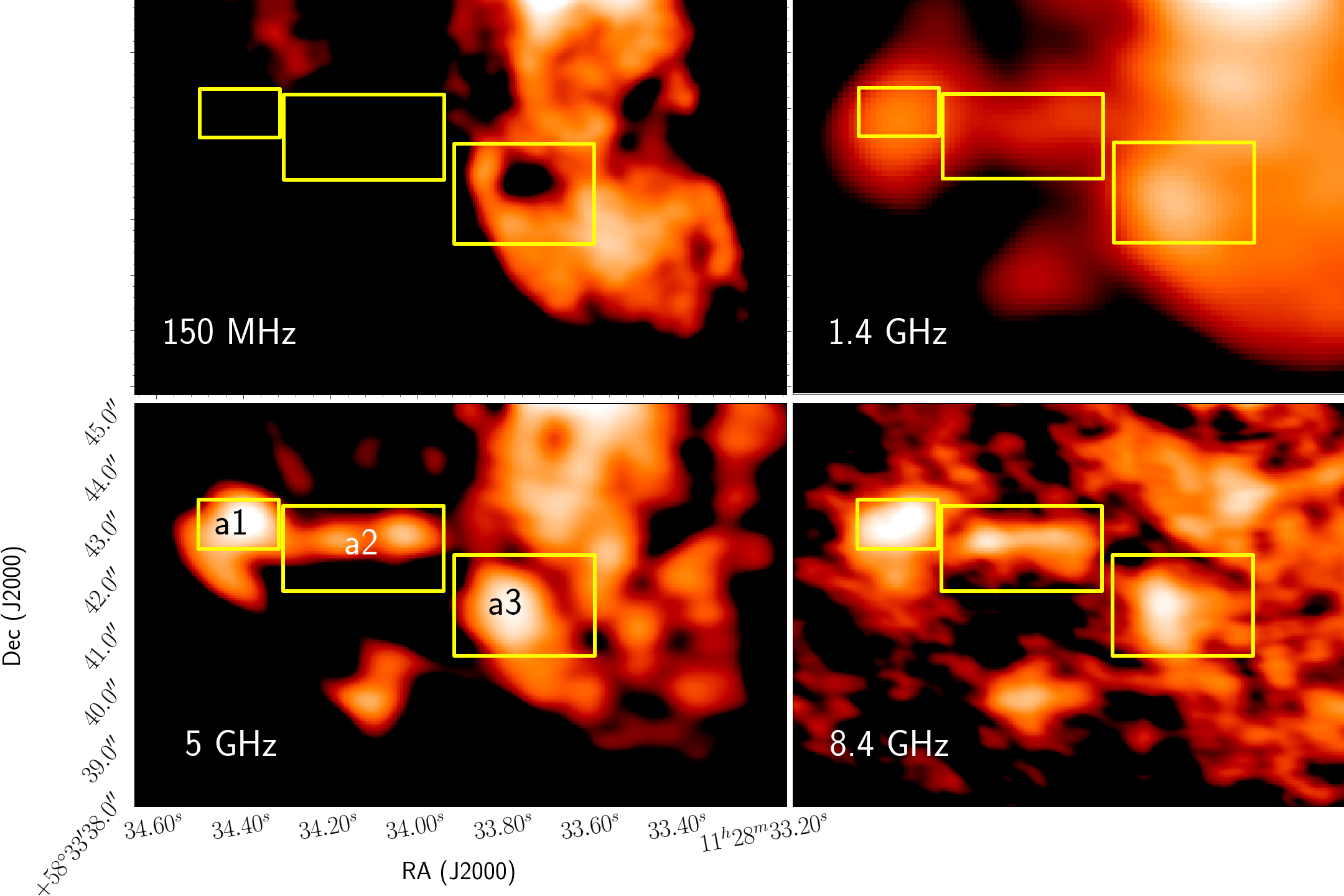}
    \caption{Detail of free/free absorption in NGC3690 East. From left to right, and top to bottom: Magnifications of the 150~MHz, and 1.4, 5, and 8.4~GHz images of Arp~299, showing the southern region of the NGC~3690-A nucleus. The different regions are  named $a_{1}$, $a_{2}$, and $a_{3}$, and are highlighted with yellow rectangles.}
    \label{fig:absorption_disk}
\end{figure*}

\begin{figure}
    \centering
    \includegraphics[width=1.0\linewidth]{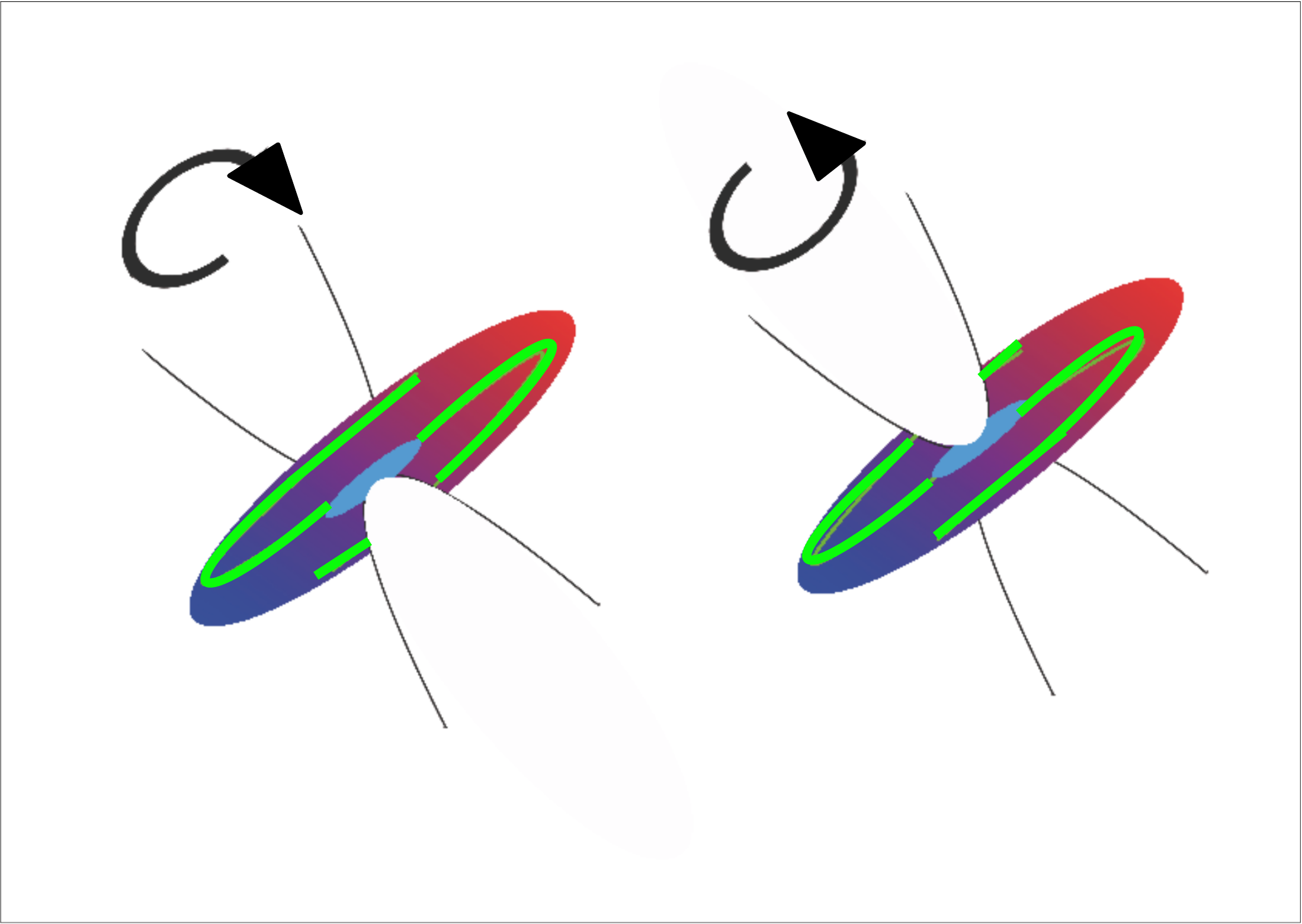}
    \caption{Sketch of the two scenarios of the disk and the outflow of NGC~3690-A. The disk is coloured in a gradient from red to blue, which represents the motion away or towards the observer, respectively. In the leading scenario (left) the observer sees the southern part of the disk, while in the  trailing scenario (right) the observer sees the northern part. The schematic shape of the spiral arms observed in the
optical and near-IR \citep[see][]{randriamanakoto} is shown in light green.}
    \label{fig:trailing}
\end{figure}

In Table \ref{tab:indirect_result} the (assumed) spectral index value of $\alpha = -0.6$ in the clumpy scenario would correspond to an injected population of relativistic electrons with $p = 2.2$ that is essentially unaffected by energy losses at our observed frequencies, or only by bremsstrahlung losses. Alternatively, the smooth continuous model leads to spectral indices for the nuclei in the range from $\alpha \simeq -0.33$ up to $\alpha -0.59$. If we neglect all energy losses, as assumed in the clumpy scenario, so that the injected spectrum is not modified, the different spectral indices would imply injected indices for the relativistic electrons of 1.6, 2.12, 2.18, and 1.78 for the A, B, C, and C' nuclei. While we cannot exclude this possibility, it seems more realistic to assume
the same injected spectrum for all nuclei. If we assume a standard injected electron index of $p = 2.1$, subject to synchrotron, bremsstrahlung, and ionization losses, the observed spectral indices of the nuclei can be explained within the continuous model by the increasing relevance of bremsstrahlung and ionization losses due the high gas densities encountered in those regions. 

\subsection{The A nucleus}

We  reported in \citet{ramirez-olivencia}, based on our 150 MHz LOFAR observations, the  detection of an outflow in the A nucleus. From energy arguments, \citet{ramirez-olivencia} concluded that the outflow was driven by the powerful nuclear starburst, where VLBI observations have detected large amounts of supernovae and supernova remnants \citep{perez-torres2009}, with a  SFR of 140 \Msunyr{} \citep{alonso-herrero2000}.

Apart from that, here we discuss another findings from our LOFAR observations. We detected emission in the  southern part of NGC~3690-A in all our JVLA images arising from three regions of $\sim$300~pc in size  ($a_{1}$, $a_{2}$, and $a_{3}$;  yellow rectangles in Fig. \ref{fig:absorption_disk}). Interestingly, the LOFAR image does not show any radio emission above the noise in these regions.  If they are star-forming regions, their non-detection at the LOFAR frequencies can be explained by free-free absorption associated with the presence of the interstellar medium ionized by young, massive stars.  Since the emission of the southern part of the outflow is absorbed at 150~MHz by
these three regions, this suggests that they belong to the disk of NGC~3690-A.

This result may have implications regarding the rotation mode of the galactic disk in NGC~3690-A, which in turn may impact the way bar formation proceeded in this galaxy in particular, and in galaxies in general \citep{kim,seo}.  
\citet{pereira-santaella} suggested, based on spectroscopic observations, 
 that the  western part of the disk of NGC~3690-A is moving away from us, while the  eastern part is moving toward us. 
 However, knowing the inclination of the disk and its Doppler rotation  is not sufficient to determine its rotation mode, as  the orientation at which we are observing the disk would still allow for two 
 possible rotation modes of the disk in NGC~3690-A (Fig. \ref{fig:trailing}): leading mode (left) and trailing mode (right).

 As mentioned earlier, we detected three free-free absorbing regions (a1, a2, and a3, Fig. \ref{fig:absorption_disk}). For a1 and a2  the absorption at 150~MHz is observed as a lack of emission, due to the free-free absorption occurring in the knot itself. For a3 the background synchrotron emission of the outflow is absorbed, as well as its own knot emission (the hole of emission in the southern part of the outflow). This indicates that we are observing the disk from the  northern part (Fig. \ref{fig:trailing}, right), and therefore the disk rotation follows the trailing mode.

\subsection{The B nucleus} \label{subsec:B_jet}

\begin{figure*}
    \centering
    \includegraphics[width=1.\linewidth]{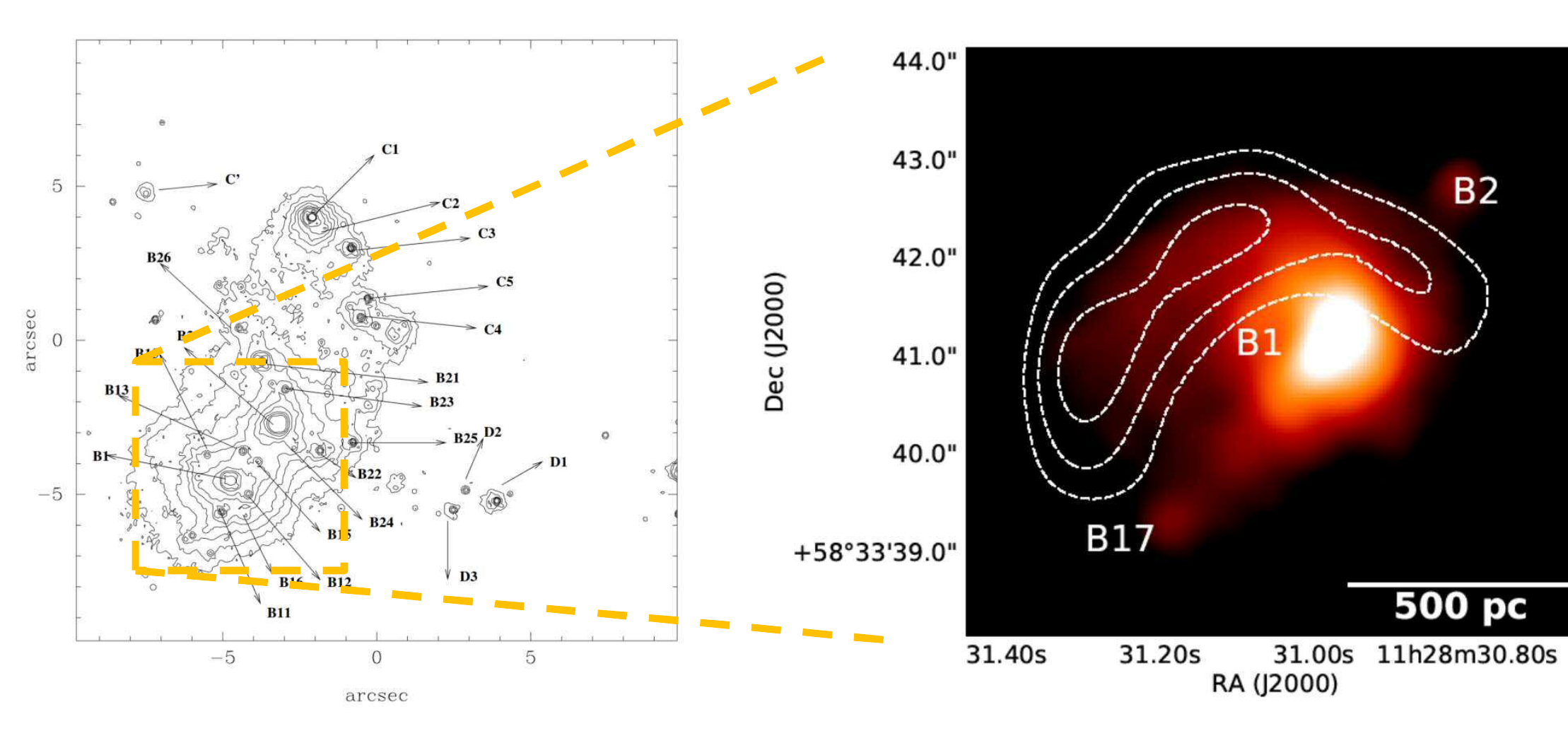}
    \caption{Detail of NGC3690 West\textit{Top panel}: NIC2 F160W image extracted from \citet{alonso-herrero2000} of the B nucleus with the names of the sources reported in it. New confirmed source B17 is indicated by a green box. \textit{Bottom panel}: Zoom-in on    the 150 MHz  image of the B nucleus, showing the positions of the nucleus (B1), a star-forming region to the north-west (B2), and B17. The white contours correspond to the spectral index between 1.4 and 5.0~GHz (-0.65, -0.7, -0.75) in the region associated with a filamentary structure, presumably a bow-shock (see main text).}
    \label{fig:bowshock_spix}
\end{figure*}

As shown in Sect.~\ref{sec:general},  the B nucleus in the LOFAR images is resolved into two main components: B1 (nucleus), the emitting region B2 to the north-west, and 
another source point-symmetric to B2 with respect to the B1 nucleus (see Fig. \ref{fig:bowshock_spix}, lower panel). We suggest that it is connected with a structure shown by \citet{alonso-herrero2000}, but not reported as a source in that work. To follow this notation we have dubbed it  B17, and it is highlighted with a green square in Fig. \ref{fig:bowshock_spix}, top panel, edited from \citet{alonso-herrero2000}. Region B17 appears to be connected to the B1 nucleus by means of a bridge of extended emission.

\subsubsection{An AGN-powered outflow in the B nucleus?}

\begin{figure}[h!]
\includegraphics[width=1.0\linewidth]{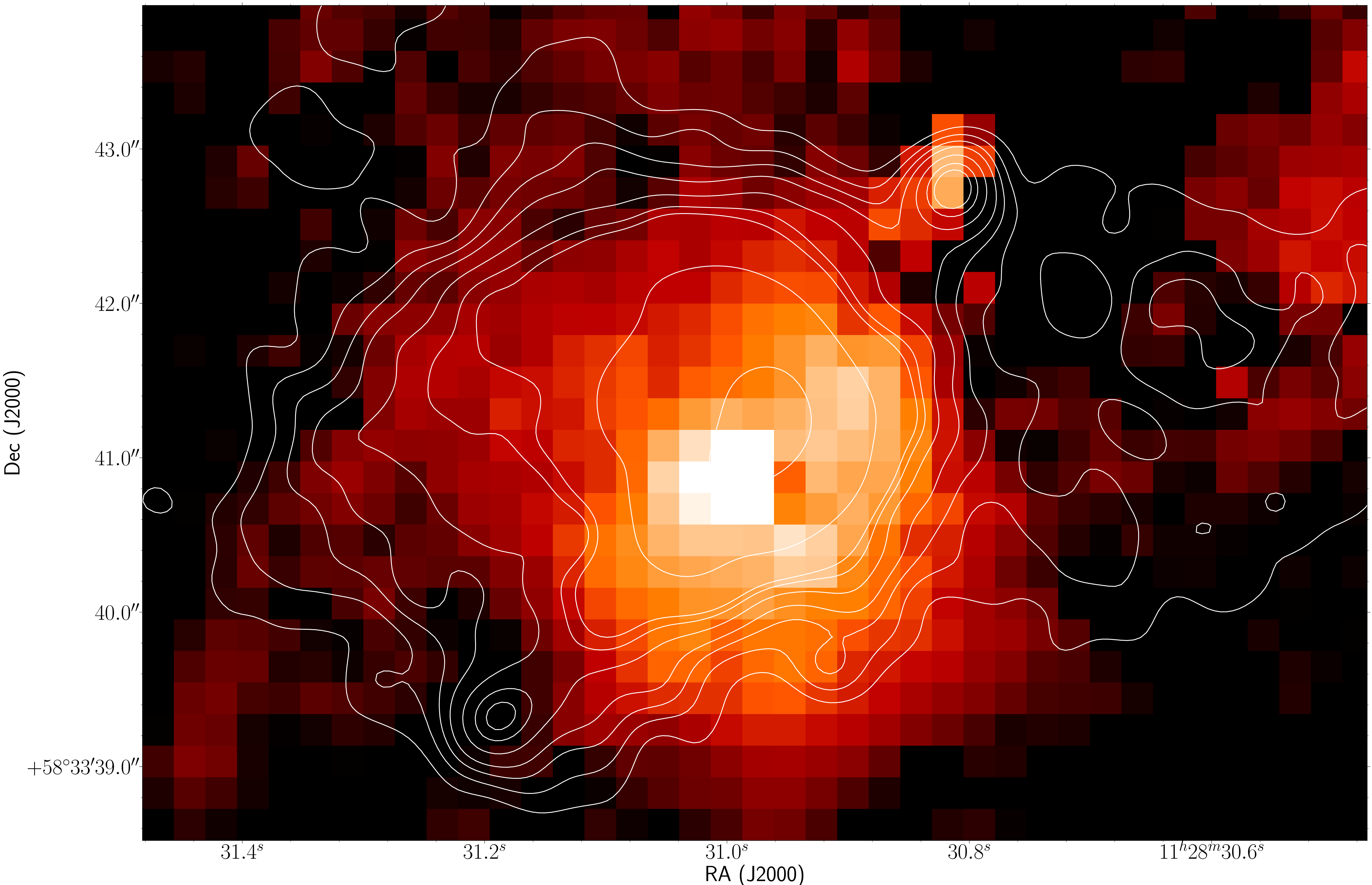}
\caption{Region B1. In  the background, continuum-subtracted NIC3 F166N, from \citet{alonso-herrero2000}, corresponding to [FeII] emission, and LOFAR at 150~MHz, in white contours.}
\label{fig:FeII-LOFAR}
\end{figure}

One of the most outstanding characteristics of the nuclear region of NGC~3690-B observed at 150~MHz is its bowshock-like structure surrounding B1 to the north, east, and west (Fig. \ref{fig:bowshock_spix}, lower panel). This bowshock could be produced by an outflow, emanating from the AGN within the B1 nucleus since it has a well-defined synchrotron behaviour (see the white contours of Fig. \ref{fig:bowshock_spix}, lower panel, with values of $\alpha_{1.4GHZ-5.0GHZ}$ of -0.6, -0.65, and -0.7). Supporting this scenario, we refer to observations at other wavelengths that trace AGN shocks. Since the extended  emission of the [FeII] infrared line is related with the shocked material due to global nuclear outflows \citep{greenhouse}, we  compare the [FeII] B1 map shown in the background in Fig. \ref{fig:FeII-LOFAR} with the LOFAR counterpart, in white contours. There appears to be weak FeII emission in the structure of the bow-shock seen in LOFAR. This [FeII] emission reinforces the scenario of an outflow or jet interacting with the surrounding medium. Another hint of this bowshock was previously presented by \citet{garcia-marin}, in which the map of the ratio  [OI]$\lambda$6300/H$\alpha$, very sensitive to shocked medium in the optical regime, presents an ionization cone emerging from B1 in the same direction as the observed bowshock. Finally, the presence of an H$_2$O maser associated with B1 also supports an outflowing scenario in this nucleus, since this line emission is known to trace shocks in the interstellar medium \citep{tarchi}. Very recently, a tidal disruption event (TDE) has been discovered \citep{mattila} in the B nucleus. We can rule out a scenario where the bowshock is  produced by the TDE, since the TDE  is very recent (the event started in 2005) and the bowshock is at least $\sim$500~pc away from the central region, so it would an intrinsic propagation speed greater than the speed of light.

\subsubsection{Nature of the possible outflow}\label{sec:beta}

Assuming the existence of an outflow in Arp~299-B1,  we studied its nature by following a similar procedure to the case of the A outflow in \citet{ramirez-olivencia} that was shown to be powered by star formation.

Let us discuss first the possibility that the outflow is AGN-powered. Considering that the luminosity of the B nucleus is $\rm L_{bol} = 3.2\times 10^{44} erg\, s^{-1}$ \citep{alonso-herrero2013}, we would obtain an accretion rate of $\dot{M}\approx0.06M_{\odot}\, yr^{-1}$.
In that case the energy rate due to the mass accretion would be
$\dot{E}_{acc_B1} \simeq 10^{11}\epsilon_{-2}\dot{M}_{acc}L_{\odot}
                 \simeq 2.34\times 10^{43} {\rm erg~s^{-1}}$, able to explain the observed outflow.
Alternatively, if the outflow is driven by star formation, we  can estimate the energy rate by assuming the core collapse supernova (CCSN) rate of $r_{SN,B1}\simeq0.3yr^{-1}$ from \citet{romero-canizales1}. Using a typical SN total energy of $E_{SN}\simeq 10^{51}erg$, and assuming an efficiency of 10\% for the conversion of energy of the explosion into mechanical energy \citep{Thor98}, we roughly obtain an energy rate from a starburst of $\dot{E}_{sb} = 9.5\cdot 10^{41}erg\, s^{-1}$, about 20 times lower than for the AGN-driven scenario. Thus, we favour the scenario in which the outflow is generated by the AGN accretion activity on the B1 nucleus, at odds with the case of the A nucleus. Moreover, the SFR in the A nucleus is SFR=140~\Msunyr{}, while in the B nucleus it is significantly lower, SFR=40~\Msunyr{}, consistent with the nature of the outflow obtained for each case.



\subsection{An intrinsic shift in the emission peaks between LOFAR and JVLA frequencies?}\label{sec:shift}

We determined the positions of the flux density peaks for the compact components at different frequencies in Sect.~\ref{sec:integrated_flux}. 
As shown in Table \ref{tab:direct_results}, the position of the peaks of the 150~MHz LOFAR image systematically differ from the position of the same peaks at all other frequencies. The position of each peak differs by less than 0.1 arcsec among all JVLA images (see Table \ref{tab:shift}), while it is much higher between the LOFAR and JVLA bands. 

\begin{figure}[h!]
\includegraphics[width=1.0\linewidth]{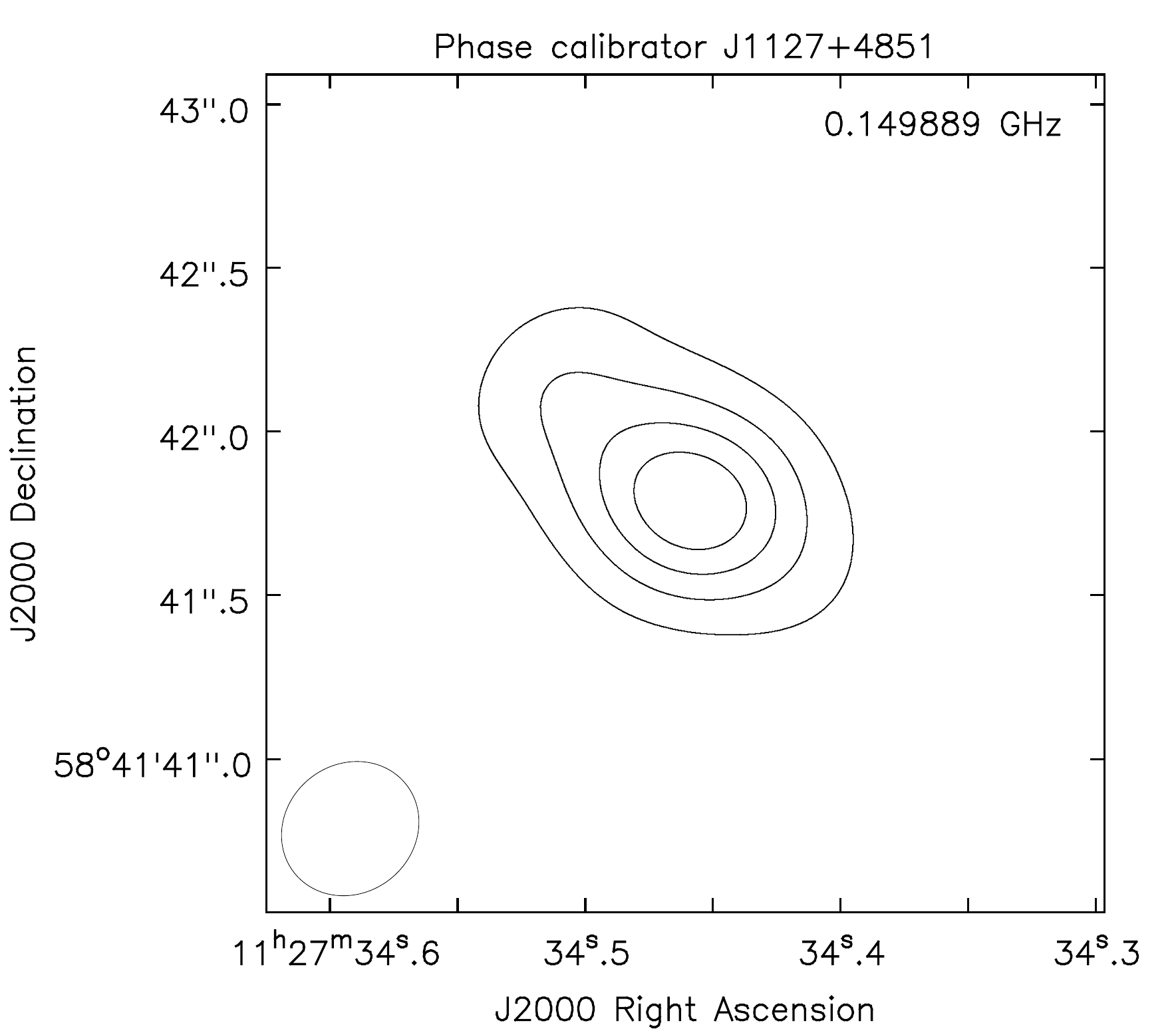}
\caption{Phase calibrator J1127+5841 with contours at [20\%, 40\%, 60\%, 80\%]$\times$ peak = 108~mJy/beam. A north-east extension is visible, with 40\% contours extending about 0.5 arcsec from the assumed catalogue position.} 
\label{fig:pcal}
\end{figure}

Differential ionospheric effects between the target and the reference fields could in principle cause apparent shifts of the target field. However, given the small separation of only 10 arcminutes,  the shifts would be negligible. 
We also considered the potential shift due to the source structure.  
We show in Fig. \ref{fig:pcal} an image of the phase calibrator for our LOFAR observations. The source displays an extended structure towards the north-east. A shift of the peak of emission in the phase-calibrator with respect to its catalogue position would produce an additional contribution to the phase solutions that propagate to the target source image. Based on the calibrator image, we estimate that the phase calibrator may contribute up to 0.5 arcsec to a (systematic) shift of all components in the north-east direction.

\begin{figure*}
    \centering
    \begin{subfigure}[b]{0.48\linewidth}        
        \centering
        \includegraphics[width=1.\linewidth]{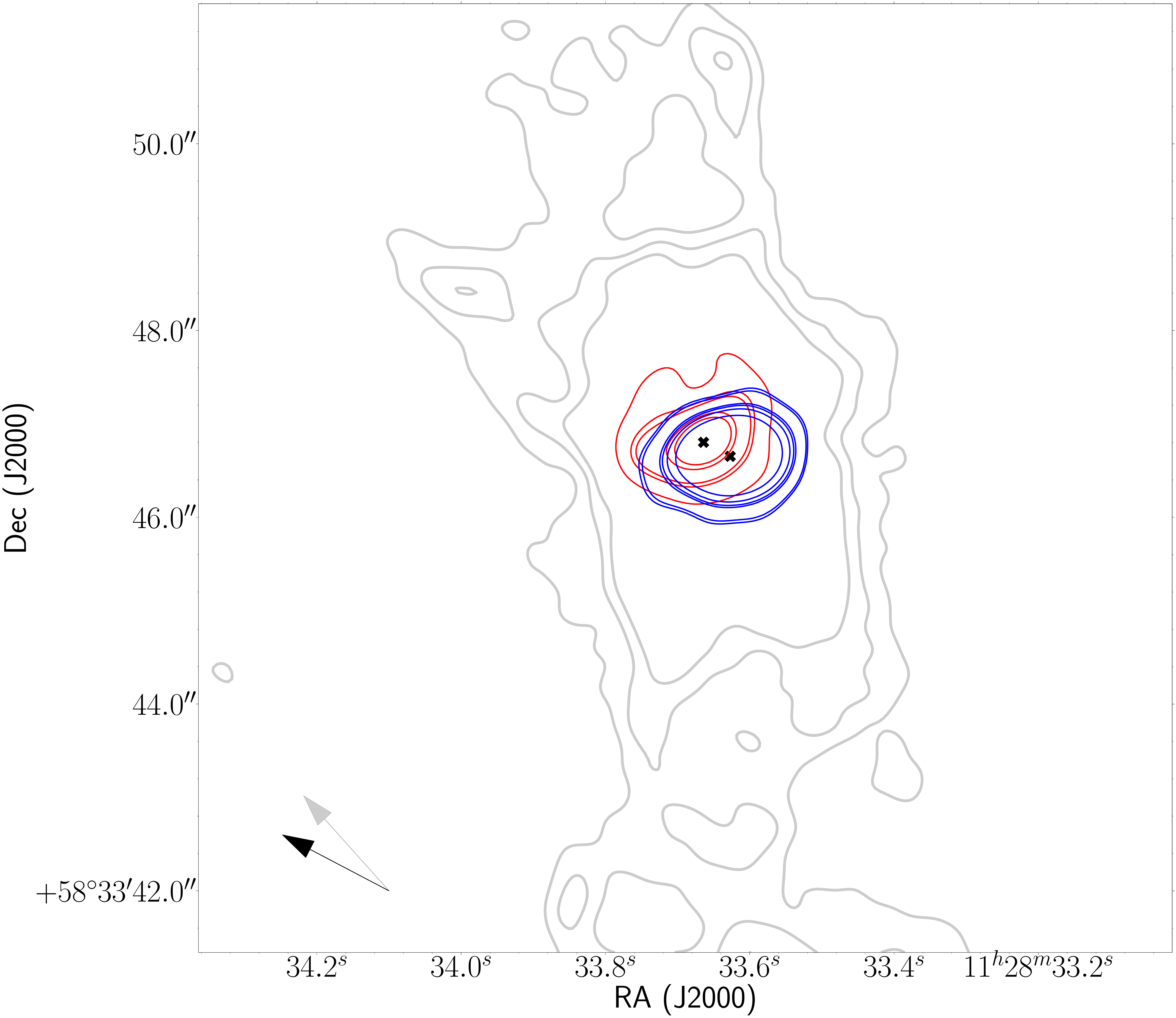}
        \caption{}
        \label{fig:shift_A}
    \end{subfigure}
    \begin{subfigure}[b]{0.48\linewidth}        
        \centering
        \includegraphics[width=1.\linewidth]{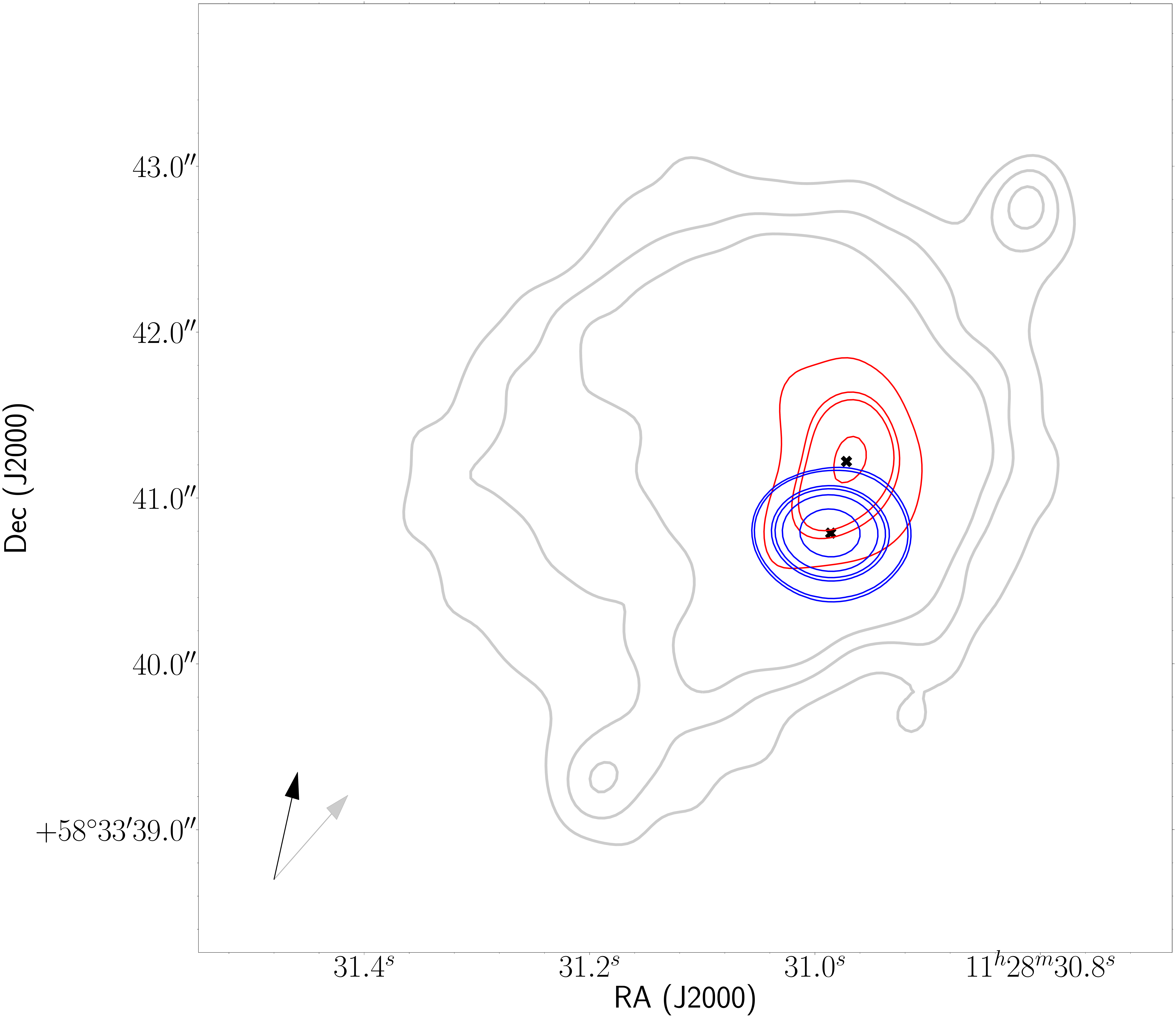}
        \caption{}
        \label{fig:shift_B}
    \end{subfigure} \\
    \begin{subfigure}[b]{0.48\linewidth}        
        \centering
        \includegraphics[width=1.\linewidth]{images/Arp299_shift_C_region.pdf}
        \caption{}
        \label{fig:shift_C}
    \end{subfigure}
    \begin{subfigure}[b]{0.48\linewidth}        
        \centering
        \includegraphics[width=1.\linewidth]{images/Arp299_shift_Cprime_region.pdf}
        \caption{}
        \label{fig:shift_Cprime}
    \end{subfigure}
    \caption{Contour plots of Arp~299-A (a),  Arp~299-B nuclei (b), and (c) C and (d) C$^{\prime}$ compact components. The red and blue contours represent the local brightest values at 150~MHz and 8.4~GHz, respectively. The grey contours in (a) and (b) show the values that shape the structures present in the two nuclei (the starburst driven outflow for A-nucleus and the assumed bowshock for B-nucleus) and detected at 150~MHz by LOFAR \citep[see][for  A-nucleus]{ramirez-olivencia}. The black crosses give the location of the maximum for each frequency. The arrows represent the directions of the shift (black) and the structure (grey).   }
    \label{fig:roc_curve}
\end{figure*}

The amplitude of the shift due to the source structure of the phase calibrator could in principle account for the maximum shift observed for the peaks of the A and B nuclei (Table~\ref{tab:shift}). 
However, this shift should be observed with the same orientation for all components, which is not the case (see the black arrows of Figs. \ref{fig:roc_curve}, which indicate different shift directions). Therefore, the existing evidence strongly suggests a physical origin for at least some of the observed shifts.
 
For the A nucleus we find that the direction of the shift is coincident with the average direction of the starburst-driven outflow (black and grey arrows, respectively, in Fig. \ref{fig:shift_A}). Similarly, we also find that the direction of the shift in the B nucleus is nearly coincident with that of its suggested outflow (Fig. \ref{fig:shift_B}).
As discussed in Sect.~\ref{sec:free-free_ISM}, high EM regions  are associated with the presence of dense ionized clouds formed at the heart of star-forming regions. Given the compact nature of the synchrotron sources (supernova remnants), the intrinsic synchrotron emission at any frequency, from LOFAR to the VLA frequencies, should come from the same location. However, the observed peaks in the LOFAR images are located at  different positions than the peaks at  higher frequencies due to the absorption by the ionized medium. The peaks observed at 1.4, 5, and 8.4~ GHz are located at the maxima of the intrinsic emission, whereas the low-frequency peaks are located where the medium starts to be barely transparent to the emission. This would naturally explain that the observed shifts in the nuclei point in different directions, as mentioned in the previous paragraph.


\begin{table}[]
\caption{Shift in the flux density peaks between pairs of frequencies for each component of Arp~299.}
\begin{tabular}{lllll}
\cline{2-5}
          & \multicolumn{4}{c}{\begin{tabular}[c]{@{}c@{}}Shift\\ {[}arcsec{]}\end{tabular}} \\ \hline
A nucleus & LOFAR             & JVLA-L             & JVLA-C             & JVLA-X             \\ \hline
LOFAR     & -                 & 0.25               & 0.31               & 0.32               \\
JVLA-L    & 0.25              & -                  & 0.06               & 0.07               \\
JVLA-C    & 0.31              & 0.06               & -                  & 0.02               \\
JVLA-X    & 0.32              & 0.07               & 0.02               & -                  \\ \hline
B nucleus &                   &                    &                    &                    \\ \hline
LOFAR     & -                 & 0.45               & 0.46               & 0.47               \\
JVLA-L    & 0.45              & -                  & 0.02               & 0.03               \\
JVLA-C    & 0.46              & 0.02               & -                  & 0.01               \\
JVLA-X    & 0.47              & 0.03               & 0.01               & -                  \\   
\hline
\end{tabular}
\label{tab:shift}
\end{table}

\subsection{C and C' regions}
At the four frequencies, the emission for  these two off-nuclear regions is extended and partially resolved, in contrast with the compact sources present in nuclei A and B. Making use of the spectral index maps (see Fig. \ref{fig:spix}) it is evident that the central areas of  nuclei C and C' have a flatter spectral index than the extended emission surrounding them, which is of synchrotron origin. This synchrotron emission would come from the star-forming regions C and C' (HII regions) resulting from the interaction between NGC~3690-A and NGC~3690-B. The flat spectral index on these physical scales  (tens to hundreds of pc) at lower frequencies (150~MHz to 1.4~GHz) can be completely explained by the absorption from H~II regions, excluding the presence of an AGN (spatial scales of a few pc). This is the case of C and C', which have a relatively flat spectral index between 1.4~GHz and 150~MHz, although they do not harbour any AGNs.

\section{Summary and conclusions}\label{sec:summary}

 We have presented the first map ever of  Arp~299 at 150 MHz, obtained from  observations with the International LOFAR telescope. We summarize our main findings as follows:

\begin{itemize}
    \item  
Our deep, subarcsecond angular resolution LOFAR observations allowed us to trace in detail the radio emission from the nuclear regions up to the  largest scales of  several kiloparsecs, at a spatial resolution better than 100 pc. The brightest components correspond to the well-known nuclei A and B (which have identified AGNs), and   to the compact components C and C'. However, most of the 150 MHz emission comes from diffuse extended structures.

\item We combined our 150 MHz LOFAR, high-resolution, deep image with deep JVLA images at 1.4, 5.0, and 8.4 GHz to obtain spatially resolved spectral index maps.  
The diffuse emission shows a two-point synchrotron spectral index between 150 MHz and 1.4 GHz, 
$\alpha \approx$ -0.7, typical of star-forming regions. In contrast, the two-point spectral index becomes flatter for the compact components.

\item We fit the radio SED of the nuclear regions using two different models of the free-free absorbing and emitting thermal gas,  distributed in  a  smooth continuous medium and a clumpy medium, respectively. 
   Both models fit  the existent data well. The smooth continuous model can explain the SED of the nuclei by a standard injected population of relativistic electrons subjected to synchrotron, bremsstrahlung, and ionization losses.
   The clumpy model can explain the data by a population of relativistic electrons with negligible energy losses, and yields thermal fractions that are more typical of star-forming galaxies compared to the continuous model. LOFAR observations at frequencies shorter than 100 MHz, or uGMRT observations at frequencies below 1 GHz would unambiguously distinguish between these two scenarios.

\item We found tentative evidence of the presence of an outflow in the B nucleus from our 150 MHz LOFAR observations. From energy considerations, this possible outflow is more likely to be AGN-driven, unlike the case of the starburst-driven outflow in Arp~299-A. 

\item We found three regions 
of about 300 pc in the A component, which are detected at high frequencies, but not in our our LOFAR image.  One of the regions is likely absorbing the southern part of the starburst-driven outflow.  This, combined with the existing Doppler information of the disk in the A component, suggests that the rotation of the disk follows a trailing mode.

\item We found a shift between the emission peaks of LOFAR and those at JVLA frequencies in the A and B  nuclei. The direction of the shift aligns well with the average direction of the outflow in the A nucleus. We also find that
the direction of the shift in the B nucleus is nearly coincident with that of its putative outflow. These shifts are not in the same direction, strongly suggesting that the shifts are of an intrinsic nature, and we suggest that differential absorption is the cause of these shifts.
 
\end{itemize}

The results obtained on Arp299 confirm the usefulness of combining spatially resolved radio imaging in the $\sim$100-200 MHz (LOFAR) up to about 1-8 GHz frequencies (JVLA)  to characterize in detail the radio emission properties of a LIRG from the central 100 pc out to the kiloparsec, galaxy-wide scales.

The SKA, which is expected to become a reality during this decade, will allow us to obtain a continuous frequency range from $\sim$100 MHz up to at least 5-8 GHz (and possibly up to around 20 GHz), so the perspectives characterizing in detail the radio emission and absorption processes in local LIRGs and ULIRGs, will be done in a systematic way.

\begin{acknowledgements}
NRO, MPT and AA acknowledge support by the Spanish MCIU through grant PGC2018-098915-B-C21 cofunded
with FEDER funds and from the State Agency for Research of the Spanish MCIU through the ``Center of Excellence Severo Ochoa'' award for the Instituto de Astrof\'{\i}sica de Andaluc\'{\i}a (SEV-2017-0709).
AA-H and MPS acknowledges support through grant PGC2018-094671-B-I00 (MCIU/AEI/FEDER,UE). AA-H  and MPS work was done under project No. MDM-2017-0737 Unidad de Excelencia “María de Maeztu”- Centro de Astrobiología (INTA-CSIC).
MPS acknowledges support from the Comunidad de Madrid, Spain, through Atracción de Talento Investigador Grant 2018-T1/TIC-11035 and PID2019-105423GA-I00 (MCIU/AEI/FEDER,UE).
\end{acknowledgements}

\bibliography{bibtex/main.bib}

\begin{appendix}
\section{Spectral models for a clumpy free-free medium} 
\label{sec:app}

We derive here an expression for the spectrum emerging from a mixed synchrotron emitting and clumpy free-free absorbing or emitting medium. This derivation is based on the clumpy medium formalism of \citet{conway}. Similar spectra are predicted  by \citet{lacki2013} who considered the detailed physical properties of the HII regions that likely comprise the free-free absorbing clumps in luminous infrared galaxies (LIRGs).  An  advantage of the  more phenomenological based derivation presented  here is that the final expression has the same free parameters as the continuous free-free model (Eq. \ref{eq:condon91}) with the addition of  $N_{cl}$ the mean number of clumps intercepted per LOS. A detailed comparison of the phenomenological and physical approaches of  respectively \citet{conway} and \citet{lacki2013} is deferred to a future paper  (Conway et al. in prep.).

For the mixed medium under consideration there are two components of emission comprising  (A) clumpy free-free emission and (B) smooth synchrotron emission mixed with a clumpy free-free absorbing medium. Although in LIRGs it is invariably the B component that dominates the radio spectrum below 10 GHz, it is convenient to consider first the A component. An expression for emission from this component is given in Eq. 15 of \citet{conway}. Converting to radio astronomy units the brightness  in  mJy/arcsec$^{2}$ is given by  
\begin{equation}
      7.22  \nu^2 (T_e/10^{4}) (1-e^{-\tau_{eff,\nu}})
\label{eq:ff_cl_emiss}
,\end{equation}
\noindent where the frequency $\nu$ is in units of GHz, $T_{e}$ is in Kelvin, and the {effective opacity} $\tau_{eff,\nu}$,  incorporating all clumping  effects, is defined by 
\begin{equation}
\tau_{eff,\nu} = N_{cl}< 1 - exp(-\tau_{cl,\nu})>
\label{eq:effopac2}
.\end{equation}

\noindent In this expression, $N_{cl}$ is the mean number of clumps intercepted per LOS, and  $\tau_{cl,\nu}$ represents the opacity at frequency $\nu$ along a LOS through an individual clump. The diagonal brackets denote an average over all impact parameters through all types of clumps within the clump  population. Here, we assume for simplicity one type of identical uniform emissivity spherical clump\footnote{\citet{lacki2013} investigated models involving a distribution of non-identical clumps (i.e. HII regions) showing only small spectral effects. Additionally, \citet{conway} shows that the exact clump shape also has little effect.}
 so the average is taken only over the impact parameter.

 The effective opacity  $\tau_{eff,\nu}$, once the clump geometry is defined, is 
  exactly determined via Eq. \ref{eq:effopac2} by the values of  $N_{cl}$ and of $<\tau_{cl,\nu}>$, the mean clump opacity averaged over all impact parameters at a frequency $\nu$.  If $\nu_{t}$ is defined as the frequency at which the {total opacity} (the mean opacity through all clumps along a LOS) equals unity, then $<\tau_{cl,\nu}>  = {N_{cl}}^{-1} (\nu/\nu_{t})^{-2.1}$.  For a spherical clump with uniform internal emissivity the opacity across a diameter (zero impact parameter) at frequency $\nu$ is $(3/2) <\tau_{cl,\nu}>$
  \citep{conway}. The value of $\tau_{cl,\nu}$ in Eq. \ref{eq:effopac2} for other impact parameters simply scales with path length through the clump. Given the above,  after taking  the average in Eq. \ref{eq:effopac2}  over all impact parameters the value of $\tau_{eff,\nu}$ versus frequency is then fully determined by the choice of $\nu_{t}$ and $N_{cl}$.
 
An expression for the  second (B) emission component, of synchrotron emission modified by clumpy free-free absorption, is given by Equation 42 of \citet{conway} as
 \begin{equation}
 \centering
   S_{\nu} 
   \frac
   { (1-e^{-\tau_{eff,\nu}})}
   {\tau_{eff,\nu}} 
\label{eq:abssync}
,\end{equation}
 
 \noindent where $S_{\nu}$ is the unabsorbed synchrotron spectrum. Following \citet{condon91} we assume that this underlying synchrotron emission is a power law with spectral index $\alpha$, linked in brightness to the free-free emission by the parameter $f_{th}$, the  thermal-to-synchrotron ratio at that frequency. The parameter $f_{th}$ is defined more precisely as the ratio of the {unabsorbed} free-free to {unabsorbed} synchrotron emission at 1 GHz. Defining the thermal-to-synchrotron ratio in this way makes it proportional to the ratio of respectively the efficiencies  of thermal emission per unit star formation and of synchrotron emission per unit star formation.
 
 The unabsorbed free-free emission at 1 GHz extrapolated from high frequency can be calculated  by taking the  high-frequency limit of
 Eq. \ref{eq:ff_cl_emiss} 
 at which the effective opacity   $\tau_{eff,\nu} << 1 $ and is equal to  $(\nu/\nu_{t})^{-2.1}$, and then setting $\nu$= 1 GHz. This gives an extrapolated unabsorbed thermal brightness at 1 GHz of  $7.22\,(T_e/10^{4})\,
 {\nu_{t}}^{2.1}$ with the unabsorbed synchrotron emission at this frequency being, by definition,  ${f_{th}}^{-1}$ times this value. Using this normalization for $S_{\nu} $ in Eq. \ref{eq:abssync} and adding the expression for the free-free clumpy emission component from Eq. \ref{eq:ff_cl_emiss},  
 the total  emerging spectrum is 
\begin{equation}
      7.22\nu^2 (T_e/10^{4})(1-e^{-\tau_{eff,\nu}})\left[1 + {\tau_{eff,\nu}}^{-1} {f_{th}}^{-1}
     {\nu_{t}}^{2.1}
     \nu^{\alpha - 2}\right]
\label{eq:clumpform}
\end{equation}
in units of  mJy/arcsec$^{2}$, where $\nu$ and $\nu_{t}$ are measured in GHz and $T_e$  in Kelvin. The effective free-free opacity  $\tau_{eff,\nu}$ in the above equation,  as described earlier, is a function of  $N_{cl}$ and $\nu_{t}$.

 It is interesting to consider Eq. \ref{eq:clumpform} in the limits of very high or very low frequencies. At high frequencies, when the opacity through individual clumps ($\tau_{cl,\nu}$) is much less than unity, $\tau_{eff,\nu} \approx N_{cl}<\tau_{\nu_cl}> = \tau_{\nu}$; in other words,  the  effective opacity is the same as the total mean opacity through all clumps along a line of sight. In this case 
  $\tau_{eff,\nu}= \left(\nu/\nu_{t}\right)^{-2.1}$, where $\nu_t$ is the frequency at which total opacity equals unity. Substituting this expression into Eq. \ref{eq:clumpform} we recover Eq. \ref{eq:condon91} which in turn describes a spectrum peaking close to frequency $\nu_t$. If we now consider very low frequencies such that individual clumps are completely optically thick,  then $\tau_{eff,\nu} \approx N_{cl}$, meaning that the effective opacity is a constant versus frequency, and thus (ignoring the very small contribution at low frequencies from free-free emission), from  Eq. \ref{eq:clumpform}, the spectrum follows the intrinsic synchrotron power-law shape after multiplication by a constant factor equal to $(1 -exp(-N_{cl}))/N_{cl}$). 
  
  The high- and low-frequency limits above give rise, for interesting cases with mean number of clumps intercepted per LOS  $ N_{cl} > 1$, to a characteristic clump model spectrum shape (see the clump models in the top two panels of Fig. \ref{fig:SED_nuclei} and  \citet{conway} Fig. 6). At  high frequencies  this spectrum is identical to that  of  the continuous model spectrum (Eq. \ref{eq:condon91})
  calculated for the same values of  $f_{th}$, $T_{e}$, $\alpha$,  and $\nu_{t}$ (i.e. a  spectrum showing a spectral peak close to frequency $\nu_{t}$). Below this turnover frequency the  spectrum at first decreases  rapidly until a local minimum is reached, below  which the spectrum turns up again. This spectral turn-up  occurs at the frequency at which  the opacity of individual clumps becomes unity; this occurs at $N_{cl}^{-1/2.1}$ times the turnover frequency,  $\nu_{t}$, at which the total opacity through the medium is unity.  In fitting the observed data, given  a low turnover frequency and a sufficiently high mean number of clumps, the spectral turn-up can occur at a frequency below that of the lowest observed frequency data point. In such cases the continuous model can provide a good fit to the data even if the free-free medium is in fact clumpy in nature. A truly continuous free-free absorbing medium is one in which the effective number of clumps goes to infinity, in which case the turn-up frequency goes to zero and the continuous free-free medium spectral model then applies at all frequencies.
  We note that \citet{conway} wrongly claimed in a footnote that the expression in \citet{condon91} leaves out a factor of $1/\tau_{\nu}$. As shown above, the general expression (Eq. \ref{eq:clumpform}),
which does include this factor, reduces to the expression of 
  \citet{condon91} in the limit of an infinite number of clumps and hence of a continuous medium.
\end{appendix}

\end{document}